\documentclass[12pt]{article}
\usepackage{jheppub}

\pdfoutput=1

\usepackage{amsmath,bbm,array,amsfonts,graphicx,wrapfig,lscape,float,mathtools,multirow,longtable}
\usepackage[dvipsnames]{xcolor}
\usepackage{array}

\newcommand{\be}{\begin{equation}}
\newcommand{\ee}{\end{equation}}
\newcommand{\beq}{\begin{equation}}
\newcommand{\beql}[1]{\begin{equation}\label{#1}}
\newcommand{\eeq}{\end{equation}}
\newcommand{\ba}{\begin{array}}
\newcommand{\ea}{\end{array}}
\newcommand{\bea}{\begin{eqnarray}}
\newcommand{\beal}[1]{\begin{eqnarray}\label{#1}}
\newcommand{\eea}{\end{eqnarray}}
\newcommand{\ben}{\begin{enumerate}}
\newcommand{\een}{\end{enumerate}}
\newcommand{\bean}{\begin{eqnarray*}}
\newcommand{\eean}{\end{eqnarray*}}
\newcommand{\eref}[1]{(\ref{#1})}
\newcommand{\sref}[1]{\S\ref{#1}}
\newcommand{\tref}[1]{Table~\ref{#1}}
\newcommand{\nn}{\nonumber}

\newcommand{\fref}[1]{Figure \ref{#1}}
\newcommand{\btab}[1]{\begin{tabular}{#1}}
\newcommand{\etab}{\end{tabular}}

\def \Black {\pdfsetcolor{0 0 0 1}}

\newcommand{\comment}[1]{}

\newcommand{\qed}{\nobreak \ifvmode \relax \else
      \ifdim\lastskip<1.5em \hskip-\lastskip
      \hskip1.5em plus0em minus0.5em \fi \nobreak
      \vrule height0.75em width0.5em depth0.25em\fi}

\usepackage{array}
\usepackage{physics}
\usepackage{subcaption}
\usepackage{tikz,tikz-3dplot}
\usepackage[colorlinks=true]{hyperref}
\usepackage{subcaption}
\newcolumntype{C}[1]{>{\centering\let\newline\\\arraybackslash\hspace{0pt}}m{#1}}

\definecolor{yellow2}{rgb}{0.98, 0.80, 0.20}

\title{Quiver-Invariant Triality \\ between Brane Brick Models} 

\author[a,b]{Dongwook Ghim,}
\author[c]{Juno Kwon,}
\author[c,d]{Rak-Kyeong Seong}

\affiliation[a]{
Particle Theory and Cosmology Group, Center for Theoretical Physics of the Universe\\
Institute for Basic Science (IBS), 55 Expo-ro, Yuseong-gu, Daejeon 34126, Korea
}
\affiliation[b]{
RIKEN Center for Interdisciplinary Theoretical and Mathematical Sciences (iTHEMS), RIKEN,
2-1 Hirosawa, Wako, Saitama 351-0198, Japan 
}

\affiliation[c]{
Department of Mathematical Sciences, and 
${}^{d}$Department of Physics,\\ 
Ulsan National Institute of Science and Technology,\\
50 UNIST-gil, Ulsan 44919, South Korea
}

\emailAdd{dongwook@ibs.re.kr}
\emailAdd{junokwon@unist.ac.kr}
\emailAdd{seong@unist.ac.kr}

\preprint{
\begin{flushright}
UNIST-MTH-26-RS-08 \\
CTPU-PTC-26-33 \\
RIKEN-iTHEMS-Report-26
\end{flushright}
}

\abstract{
We study quiver-invariant triality between $2d$ $(0,2)$ supersymmetric gauge theories described by brane brick models associated with toric Calabi-Yau 4-folds.
The dual theories share the same gauge groups, field content and $E$-terms but have distinct $J$-terms.
We present an explicit pair of such brane brick models and identify the sequence of triality mutations relating them.
This sequence returns the quiver to its original form while transforming the $J$- and $E$-terms.
We explain the relation between quiver-invariant triality and the orbifold reduction of quiver-invariant brane tilings.
}

\begin{document}

\maketitle

\section{Introduction \label{sec01}}

Supersymmetric gauge theories on D-branes probing toric Calabi-Yau singularities provide a natural setting in which to study the interplay between gauge theory and geometry.
For D3-branes probing toric Calabi-Yau 3-folds, brane tilings encode the matter content and superpotential of the corresponding $4d$ $\mathcal{N}=1$ theories \cite{Hanany:2005ve,Franco:2005rj,Franco:2005sm,Kennaway:2007tq,Yamazaki:2008bt}.
Brane brick models extend this description to the $2d$ $(0,2)$ theories on D1-branes probing toric Calabi-Yau 4-folds \cite{Franco:2015tna,Franco:2015tya}.
Their dual periodic quivers on a 3-torus $T^3$ represent chiral and Fermi fields by directed and undirected edges, respectively,
and encode the $J$- and $E$-terms through interaction plaquettes.

The underlying quiver specifies the gauge groups and matter content, but does not determine the $J$- and $E$-terms.
This distinction is particularly relevant when comparing brane brick models associated with the same toric Calabi-Yau 4-fold.
Gadde-Gukov-Putrov triality relates $2d$ $(0,2)$ theories that are equivalent in the IR \cite{Gadde:2013lxa}.
For brane brick models, triality acts locally on the periodic quiver on $T^3$ and preserves the mesonic moduli space \cite{Franco:2016nwv,Franco:2016qxh}.
Although an individual triality mutation generally changes the underlying quiver, we show that a sequence of such mutations
can return it to its original form while changing the $J$- and $E$-terms and the associated periodic quiver.
We refer to this phenomenon as \textit{quiver-invariant triality}.

A related phenomenon was observed for brane tilings in \cite{Kho:2026geo,Lee:2026tju}.
For brane tilings, $4d$ $\mathcal{N}=1$ theories with the same quiver and mesonic moduli space but distinct superpotentials are connected by sequences of Seiberg dualities \cite{Seiberg:1994pq,Kho:2026geo,Lee:2026tju}.
Dimensional reduction, or more generally orbifold reduction \cite{Franco:2016fxm,Franco:2015tna,Franco:2015tya}, provides a way to construct brane brick models from brane tilings.
We use orbifold reduction to obtain examples of quiver-invariant triality from brane tilings related by quiver-invariant dualities.

In this work, we study a pair of brane brick models that share the same quiver with $18$ gauge nodes, $56$ chiral fields and
$38$ Fermi fields, and have identical $E$-terms but distinct $J$-terms.
Using the forward algorithm \cite{Franco:2015tna,Franco:2015tya,Feng:2000mi}, we show that the two brane brick models have the same mesonic moduli space and hence correspond to the same toric Calabi-Yau 4-fold.
We compute the Hilbert series \cite{Benvenuti:2006qr,Feng:2007ur,Butti:2007jv,Hanany:2006uc,Hanany:2007zz,Franco:2015tya} of this common mesonic moduli space refined in terms of fugacities for the enhanced
mesonic flavor symmetry $SU(2)_x\times U(1)_{f_1}\times U(1)_{f_2}$.

The paper is organized as follows. Section \sref{sec02} reviews brane brick models, triality and orbifold reduction.
Section \sref{sec03} presents the two brane brick models related by quiver-invariant triality, their common mesonic moduli space,
their elliptic genus and the triality sequence connecting them.
We conclude in section \sref{sec05} with a discussion of our results and future directions.
\\

\section{Background \label{sec02}}

In the following section, we review $2d$ $(0,2)$ quiver gauge theories realized as worldvolume theories on D1-branes probing toric Calabi-Yau 4-folds
and their description in terms of brane brick models \cite{Franco:2015tna,Franco:2015tya}.
We then summarize
Gadde-Gukov-Putrov triality \cite{Gadde:2013lxa,Franco:2016nwv} and its realization as a local mutation of brane brick models.
Finally, 
the section reviews $4d$ $\mathcal{N}=1$ quiver gauge theories realized in terms of brane tilings \cite{Hanany:2005ve,Franco:2005rj,Franco:2005sm,Kennaway:2007tq,Yamazaki:2008bt} 
and how they can be used to obtain brane brick models via a method known as orbifold reduction \cite{Franco:2016fxm}.
\\

\subsection{$2d$ $(0,2)$ Quiver Gauge Theory and Brane Brick Models \label{sec021}}

\paragraph{Brane brick models and $2d$ $(0,2)$ gauge theories.}
The worldvolume theories on 
$N$ D1-branes probing a toric Calabi-Yau 4-fold form 
a rich family of $2d$ $(0,2)$ quiver gauge theories with $U(N)$ gauge groups.
Under T-duality, the probe D1-branes become D4-branes
suspended between an NS5-brane wrapping a holomorphic surface $\Sigma$ \cite{Franco:2015tna,Franco:2015tya} associated with the toric Calabi-Yau 4-fold.
The toric Calabi-Yau 4-fold is characterized by a convex lattice
polytope $\Delta\subset\mathbb{Z}^3$ known as its toric diagram \cite{fulton1993introduction,CLS}.
The associated Newton polynomial is given by, 
\beal{es01a00}
P(x,y,z) = 
\sum_{(n_x,n_y,n_z) \in \Delta} ~c_{(n_x,n_y,n_z)}~
x^{n_x} y^{n_y} z^{n_z} ~,~
\eea
where $x,y,z\in \mathbb{C}^*$ and the arguments of the complex coordinates $x,y,z$ parameterize a 3-torus $T^3$.
The holomorphic surface $\Sigma$ is then given by the zero locus of the Newton polynomial, 
\beal{es01a01}
\Sigma ~:~ P(x,y,z) = 0 ~.~
\eea

\begin{figure}[http!!]
\centering
\includegraphics[height=0.6\textheight]{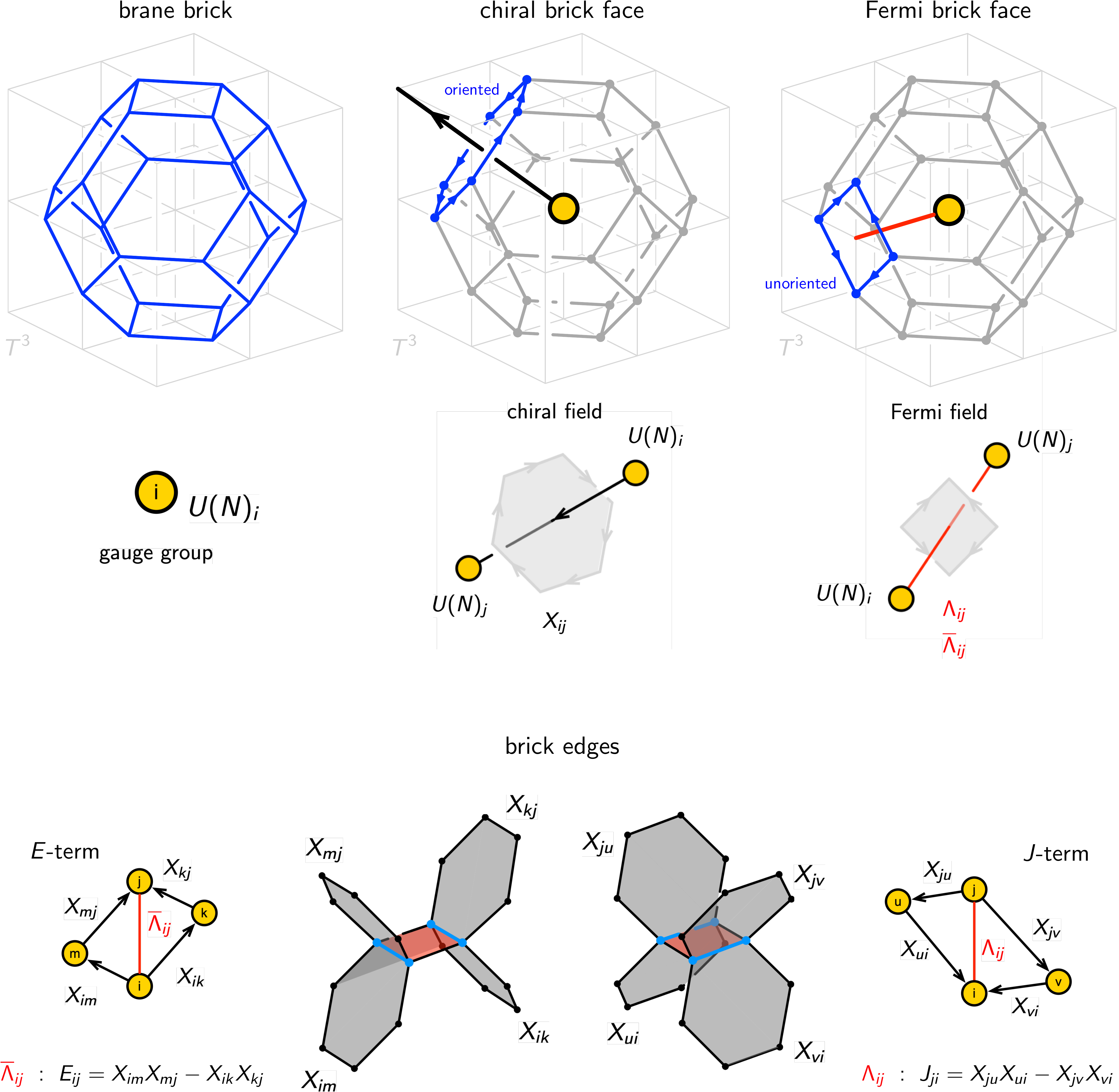}
\caption{The dictionary between brane brick models and $2d$ $(0,2)$ gauge theories.}
\label{fig_01}
\end{figure}

The coamoeba of $\Sigma$ is its image under the argument map to the 3-torus $T^3$ \cite{Franco:2015tya,Franco:2015tna,Futaki:2014mpa}.
In the corresponding tropical limit, 
the NS5-brane is represented by polygonal faces that divide $T^3$ into 3-dimensional polytopes filled by D4-branes.
This tessellation, together with the encoded gauge theory data, is called a brane brick model \cite{Franco:2015tna,Franco:2015tya}.
As summarized in the dictionary first introduced in \cite{Franco:2015tna,Franco:2015tya} and illustrated in \fref{fig_01},
the polytopes in the tessellation of $T^3$ correspond to $U(N)$ gauge groups of the associated $2d$ $(0,2)$ theory 
and are referred to as brane bricks.
The polygonal boundary faces of the brane bricks correspond to either bifundamental chiral fields $X_{ij}$
or Fermi fields $\Lambda_{ij}$
adjacent to a pair of brane bricks associated with $U(N)_i \times U(N)_j$.
The edges of the tessellation correspond to plaquettes in the periodic quiver, which are formed by a monomial from a $J$- or $E$-term together with the corresponding Fermi field or its conjugate \cite{Franco:2015tya}.

Brane brick models are in one-to-one correspondence with periodic quivers
that encode both the quiver and the $J$- and $E$-terms of the $2d$ $(0,2)$ theory.
Periodic quivers are obtained by graph dualization of the brane brick model on $T^3$.
Throughout this work,
we make use of periodic quivers to identify and compare different $2d$ $(0,2)$ theories realized by brane brick models.
\\

\paragraph{Quivers and $J$- and $E$-terms.}
The nodes of the quiver of a brane brick model 
correspond to $U(N)_i$ gauge groups in the corresponding $2d$ $(0,2)$ theory.
The bifundamental chiral multiplets $X_{ij}$ are given by directed edges in the quiver 
whereas Fermi multiplets $\Lambda_{ij}$ and their associated conjugates $\bar{\Lambda}_{ij}$ are represented by undirected edges. 
The undirected edges reflect the invariance of the $2d$ $(0,2)$ theory under $\Lambda_{ij}\leftrightarrow\bar{\Lambda}_{ij}$ accompanied by the exchange of the corresponding $J$- and $E$-terms \cite{Franco:2015tna,Franco:2015tya}.

For every Fermi field $\Lambda_{ij}$, there is a pair of $J_{ji}$- and $E_{ij}$-terms.
The $J_{ji}$- and $E_{ij}$-terms associated to each $\Lambda_{ij}$ are constrained by $(0,2)$ supersymmetry and 
accordingly satisfy the following trace condition \cite{Franco:2015tna,Franco:2017cjj}, 
\beal{es01a05}
\sum_{\Lambda_{ij}} \tr(E_{ij} J_{ji}) = 0~,~
\eea
where the sum runs over all Fermi multiplets $\Lambda_{ij}$ in the $2d$ $(0,2)$ theory.
For brane brick models, the $J_{ji}$- and $E_{ij}$-terms are binomial under the toric condition \cite{Franco:2015tna,Franco:2015tya}.
They take the following general form, 
\beal{es01a06}
J_{ji} = J_{ji}^+ - J_{ji}^- ~,~
E_{ij} = E_{ij}^+ - E_{ij}^- ~,~
\eea
where $J_{ji}^\pm$ and $E_{ij}^\pm$ are monomial products of chiral fields.
The products $\Lambda_{ij}J_{ji}^\pm$ and $\bar{\Lambda}_{ij}E_{ij}^\pm$ form closed paths in the periodic quiver and are referred to as plaquettes.

For equal gauge group ranks, cancellation of the non-abelian gauge anomalies \cite{Franco:2015tna,Franco:2015tya} requires the following condition at every quiver node,
\beal{es01a07}
n_i^\chi - n_i^F = 2 ~,~
\eea
where $n_i^\chi$ and $n_i^F$ count the chiral and Fermi fields attached to node $i$, respectively, with each adjoint field counted twice.
Summing over the nodes of the quiver gives the relation, 
\beal{es01a08}
n^\chi = G + n^F ~,~
\eea
where $n^\chi$ and $n^F$
are the total number of chiral and Fermi fields, respectively, 
and $G$ is the total number of $U(N)$ gauge groups in the $2d$ $(0,2)$ theory.
\\

\paragraph{Mesonic moduli space.}
For abelian $2d$ $(0,2)$ gauge theories
with $U(1)$ gauge groups, 
the \textit{mesonic moduli space} is defined as follows,
\beal{es01a10}
\mathcal{M}^{mes} = \text{Spec} (\mathbb{C}[X_{ij}] / \mathcal{I}_{Irr}) // U(1)^{G-1} ~,~
\eea
where the coordinate ring $\mathbb{C} [X_{ij}]$
is defined in terms of the set of bifundamental chiral fields $X_{ij}$, 
and $\mathcal{I}_{Irr}$ is the ideal defining the coherent component of the variety determined by the vanishing $J$- and $E$-terms.
The diagonal $U(1)$ decouples, leaving a quotient by $U(1)^{G-1}$.
The coordinate ring of the mesonic moduli space consists of mesonic gauge-invariant chiral operators subject to these relations.
For an abelian brane brick model, this moduli space is the associated toric Calabi-Yau 4-fold.

We can express the mesonic moduli space in terms of GLSM fields $p_a$ \cite{Witten:1993yc}, with $a=1, \dots, c$.
Among the GLSM fields $p_a$, those having a combinatorial interpretation correspond to brick matchings, 
which are subsets of bifundamental chiral and Fermi fields defined combinatorially in \cite{Franco:2015tya}.
GLSM fields corresponding to brick matchings \cite{Franco:2015tya} are associated with lattice points in the toric diagram.
Those associated with extremal vertices are referred to as extremal GLSM fields.
Here, we express the chiral fields in terms of products of GLSM fields $p_a$ as follows,
\beal{es01a11}
X_{k} = \prod_{a=1}^c ~p_a^{P_{ka}} ~,~
\eea
where the $P$-matrix is given by,
\beal{es01a12}
P_{ka} =
\left\{
\ba{ll}
1 & \text{~~if $X_k \in p_a$}\\
0 & \text{~~if $X_k \notin p_a$}
\ea
\right.
~.~
\eea
Here, we note that 
columns in the $P$-matrix corresponding to extra GLSM fields obtained from the
forward algorithm can have entries larger than $1$.

The $P$-matrix can be also obtained via the forward algorithm \cite{Franco:2015tna,Franco:2015tya,Feng:2000mi}
from the $J$- and $E$-terms of the brane brick model.
In terms of the GLSM fields $p_a$ defined by the $P$-matrix, we can express
the $J$- and $E$-terms as well as the $D$-terms of the brane brick model 
as $U(1)$ charges on the GLSM fields given by
the $Q_{JE}$ and $Q_D$ matrices, respectively.
They are defined as follows,
\beal{es01a15}
Q_{JE} = \ker (P)
~,~ \quad
\overline{d} = Q_D \cdot P^{t}
~,~
\eea
where $\overline{d}$ is the reduced incidence matrix for chiral fields in the quiver, 
which has $G-1$ rows.

In terms of the GLSM fields $p_a$
and the $U(1)$ charges given by the $Q_{JE}$ and $Q_D$ matrices, 
the mesonic moduli space can be expressed in terms of the following symplectic quotient,
\beal{es01a02}
\mathcal{M}^{mes} = (\text{Spec} ~ \mathbb{C}[p_a] ~ // Q_{JE} ) // Q_{D} ~.~
\eea
As noted above, the vertices of the toric diagram of the toric Calabi-Yau 4-fold
are associated with the GLSM fields that parameterize the mesonic moduli space as given in \eref{es01a02}.
The coordinates of the vertices and hence the toric diagram of the toric Calabi-Yau 4-fold are given by,
\beal{es01a20}
G_t = \text{ker}(Q_t) ~,~
\eea
where the total charge matrix $Q_t$ is defined as a concatenation of $Q_{JE}$ and $Q_D$.
Here, the columns of the $G_t$-matrix are associated with the coordinates of the vertices $v_a$
of the toric diagram $\Delta$ as well as the corresponding GLSM fields $p_a$.
\\

\paragraph{Hilbert series.}
The generating function of mesonic gauge invariant operators of $\mathcal{M}^{mes}$
is given by the corresponding Hilbert series \cite{Benvenuti:2006qr,Feng:2007ur,Butti:2007jv,Hanany:2006uc,Hanany:2007zz}. 
By expressing the mesonic gauge invariant operators in terms of GLSM fields $p_a$, 
we define the Hilbert series of $\mathcal{M}^{mes}$ in an abelian brane brick model as follows,
\beal{es01a30}
g(t_a; \mathcal{M}^{mes}) = 
\prod_{m=1}^{c-4} \oint_{|z_m|=1} \frac{d z_m}{2\pi i z_m} \prod_{a=1}^{c}
\frac{1}{1-t_a \prod_{n=1}^{c-4}z_n^{(Q_t)_{na}}}
~,~
\eea
where $c$ is the number of GLSM fields and $t_a$ is the fugacity counting the degree in $p_a$.
Using the plethystic program \cite{Benvenuti:2006qr,Feng:2007ur,Hanany:2007zz,Butti:2007jv,Hanany:2006uc}, the Hilbert series also provides information about the generators and relations of
$\mathcal{M}^{mes}$, including their degrees and global symmetry representations.
\\

\paragraph{Elliptic genus.}
The elliptic genus is a supersymmetric index of a $2d$ $(0,2)$ theory, defined as a twisted partition function on the torus $T^2$.
It can be computed from the $2d$ $(0,2)$ gauge theory by supersymmetric localization, which reduces the path integral to a contour integral evaluated using the Jeffrey-Kirwan residue prescription \cite{Benini:2013nda,Benini:2013xpa,Franco:2017cjj}.
For brane brick models with a large number of gauge nodes, however, the direct evaluation of these residues becomes computationally challenging. 

For abelian brane brick models, an alternative approach is provided by the geometric formula for the modified elliptic genus introduced in \cite{Franco:2017cjj}.
This formula expresses the index in terms of a triangulation of the toric diagram of the associated Calabi-Yau 4-fold.
In examples where both methods can be applied, the geometric computation agrees with the gauge theory localization result.
Let $\Delta$ be the toric diagram of the toric Calabi-Yau 4-fold admitting a triangulation $\mathcal{T}$ into unimodular tetrahedra.
The geometric pre-index is then given by,
\beal{es01a31}
\mathcal{J}_{\Delta}(\mathbf{t})
=
-i\eta(q)^3
\sum_{\sigma\in\mathcal{T}}
\frac{
\prod_{a=1}^{3}\theta_1\bigl(q,z_a^{\sigma}(\mathbf{t})\bigr)
}
{
\prod_{b=1}^{4}\theta_1\bigl(q,y_b^{\sigma}(\mathbf{t})\bigr)
}
~.~
\eea
Here, $\mathbf{t}=(t_1,t_2,t_3,t_4)$ denotes the geometric fugacities
associated with the toric diagram $\Delta$.
The Dedekind eta function and Jacobi theta function, written in
multiplicative notation, 
are defined as follows,
\beal{es01a32}
\eta(q)
&=&
q^{1/24}\prod_{n=1}^{\infty}(1-q^n)~,~
\nn\\
\theta_1(q,y)
&=&
-iq^{1/8}y^{1/2}
\prod_{n=1}^{\infty}(1-q^n)(1-yq^n)(1-y^{-1}q^{n-1})
~.~
\eea

For each tetrahedron $\sigma\in\mathcal{T}$, the arguments of the theta functions in the denominator of \eref{es01a31} are given by,
\beal{es01a33}
y_a^{\sigma}(\mathbf{t})
=
\prod_{i=1}^{4}~t_i^{(w_a^{\sigma})_i}
~,~
 a=1,\ldots,4
 ~,~
\eea
where $w_a^{\sigma}$ are the dual vectors associated with the $4$ vertices $v_{\sigma_a}\in\mathbb{Z}^4$ of $\sigma$.
For an oriented unimodular tetrahedron, they satisfy $v_{\sigma_b}^i(w_a^{\sigma})_i=\delta_{ab}$ and can be written as follows,
\beal{es01a34}
(w_a^{\sigma})_i
=\frac{1}{3!}\,
\epsilon^{abcd}\epsilon_{ijkl}
~
v_{\sigma_b}^{j}v_{\sigma_c}^{k}v_{\sigma_d}^{l}
 ~.~
\eea
Here, 
the vertices are ordered so that
the determinant of $(v_{\sigma_1},\ldots,v_{\sigma_4})$ is $+1$.
The arguments of the theta functions in the numerator are given by,
\beal{es01a35}
z_1^{\sigma}(\mathbf{t})
=\sqrt{\frac{y_1^{\sigma}y_4^{\sigma}}{y_2^{\sigma}y_3^{\sigma}}}
~,~
z_2^{\sigma}(\mathbf{t})
=\sqrt{\frac{y_2^{\sigma}y_4^{\sigma}}{y_1^{\sigma}y_3^{\sigma}}}
~,~
z_3^{\sigma}(\mathbf{t})
=\sqrt{\frac{y_3^{\sigma}y_4^{\sigma}}{y_1^{\sigma}y_2^{\sigma}}}
~.~
\eea

The modified elliptic genus is obtained by expressing the geometric fugacities in terms of the three mesonic flavor fugacities
$\mathbf{x}=(x_1,x_2,x_3)=(x,f_1,f_2)$ and setting the $U(1)_R$ fugacity to $1$
giving us,
\beal{es01a36}
\mathcal{I}_{\Delta}(q;\mathbf{x})
=\left.\mathcal{J}_{\Delta}(\mathbf{t})\right|_{
 t_i=\prod_{k=1}^{3}x_k^{m_{ik}}}
 ~,~
i=1,\ldots,4
~.~
\eea
The exponents $m_{ik}$ are determined by matching the
mesonic flavor charges of gauge invariant operators in the mesonic moduli space with their counterparts in the toric geometry,
as encoded in the Hilbert series refined by the mesonic global symmetry fugacities \cite{Franco:2017cjj}.
\\

\subsection{Triality between $2d$ $(0,2)$ Theories and Brane Brick Models \label{sec022}}

\begin{figure}[http!!]
\centering
\includegraphics[height=0.35\textheight]{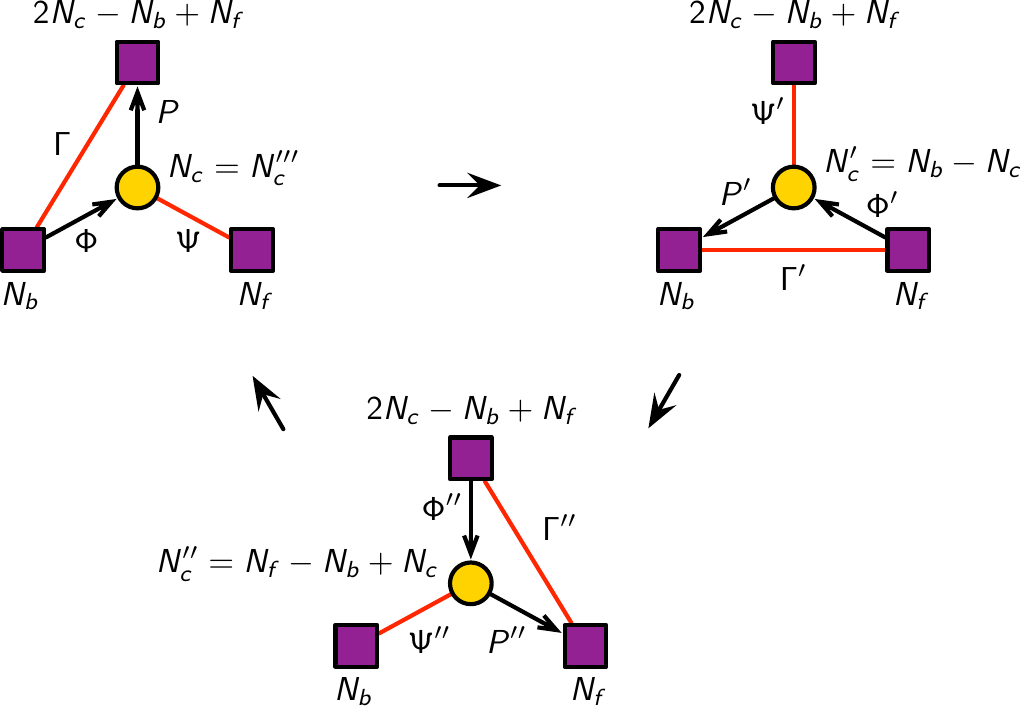}
\caption{The triality loop for $2d$ $(0,2)$ SQCD \cite{Gadde:2013lxa,Franco:2016nwv}.}
\label{fig_03}
\end{figure}

\paragraph{Triality on Brane Brick Models.}
Triality between $2d$ $(0,2)$ supersymmetric gauge theories
was first observed and studied by Gadde, Gukov and Putrov in \cite{Gadde:2013lxa}.
In its original formulation, the $U(N_c)$ theory contains 
$N_b$ fundamental chiral fields $\Phi$,
$N_f$ anti-fundamental Fermi fields $\Psi$,
$2N_c-N_b+N_f$ anti-fundamental chiral fields $P$,
and Fermi fields $\Gamma$ with $J_\Gamma=\Phi P$,
where flavor indices are suppressed.
Triality relates this $2d$ $(0,2)$ theory to other theories
with gauge group ranks
given by, 
\beal{es01a40}
N_c^\prime = N_b-N_c ~,~
N_c^{\prime\prime} = N_f - N_b + N_c ~,~
N_c^{\prime\prime\prime} = N_c ~,~
\eea
where the numbers of flavors are permuted cyclically at each step of the triality transformation as illustrated in \fref{fig_03}.
The original theory is recovered after $3$ consecutive triality transformations at the same gauge group.

For $2d$ $(0,2)$ quiver gauge theories
corresponding to brane brick models, 
triality has a brane realization as first introduced in \cite{Franco:2016nwv,Franco:2016qxh}.
For brane brick models with $U(N)$ gauge groups, 
the rank of the dualized gauge node given by $U(N_k)$ 
becomes under triality, 
\beal{es01a42}
N_k^\prime = \sum_{j \neq k} n_{jk}^\chi N_j - N_k ~,~
\eea 
where $n_{jk}^{\chi}$ is the number of chiral fields from gauge node $j$ to node $k$ in the quiver.
Additionally, the chiral and Fermi fields in the quiver transform relative to $U(N_k)$ as follows, 
\begin{itemize}
\item 
$X_{mk} \rightarrow X_{km}^\prime$, $X_{km} \rightarrow \Lambda_{km}^\prime$, $\Lambda_{mk} \rightarrow X_{mk}^\prime$
\item
$X_{mk} X_{kn} \rightarrow X_{mn}^\prime$
\item
$X_{mk}\Lambda_{kn} \rightarrow \Lambda_{mn}^\prime$
\item
remove every massive chiral-Fermi pair $(X_{mn}^\prime, \Lambda_{mn}^\prime)$
\end{itemize}
after which the $J$- and $E$-terms are identified from the resulting periodic quiver of the brane brick model.
Let $\tau_k$ denote triality at gauge node $k$, and let $\gamma$
denote chiral conjugation, which reverses all chiral arrows
and exchanges the $J$- and $E$-terms.
These operations satisfy the following constraints, 
\beal{s01a43}
\tau_k^3 = 1 ~,~
\gamma^2 = 1 ~,~
\gamma \cdot \tau_k \cdot \gamma = \tau_k^{-1} = \tau_k^2 ~,~
\eea
where $\tau_k^2$ represents the inverse triality operation at a gauge node $k$.
These transformations at different gauge nodes of the $2d$ $(0,2)$ quiver
generate a triality network as first presented for various examples in \cite{Gadde:2013lxa,Franco:2016nwv}.

Not every gauge node may be dualized without 
leaving the family of $2d$ $(0,2)$ theories described by brane brick models.
Preserving equal gauge ranks in \eref{es01a42} requires $n_{k,\text{in}}^\chi=2$, while anomaly cancellation implies $n_{k,\text{out}}^\chi=n_k^\Lambda$.
Thus, the dualized node must satisfy, 
\beal{es01a45}
n_{k, \text{in}}^\chi = 2 ~,~
n_{k, \text{out}}^\chi = n_k^\Lambda \geq 2 ~,~
\eea
where $n_{k, \text{in}}^\chi$ and $n_{k, \text{out}}^\chi$
are the number of incoming and outgoing chiral field arrows,
and $n_k^\Lambda$ is the number of Fermi fields at gauge node $k$ in the quiver.
For admissible local configurations of the periodic quiver, triality at such nodes produces another toric phase associated
with the same toric Calabi-Yau 4-fold.

\begin{figure}[http!!]
\centering
\includegraphics[height=0.38\textheight]{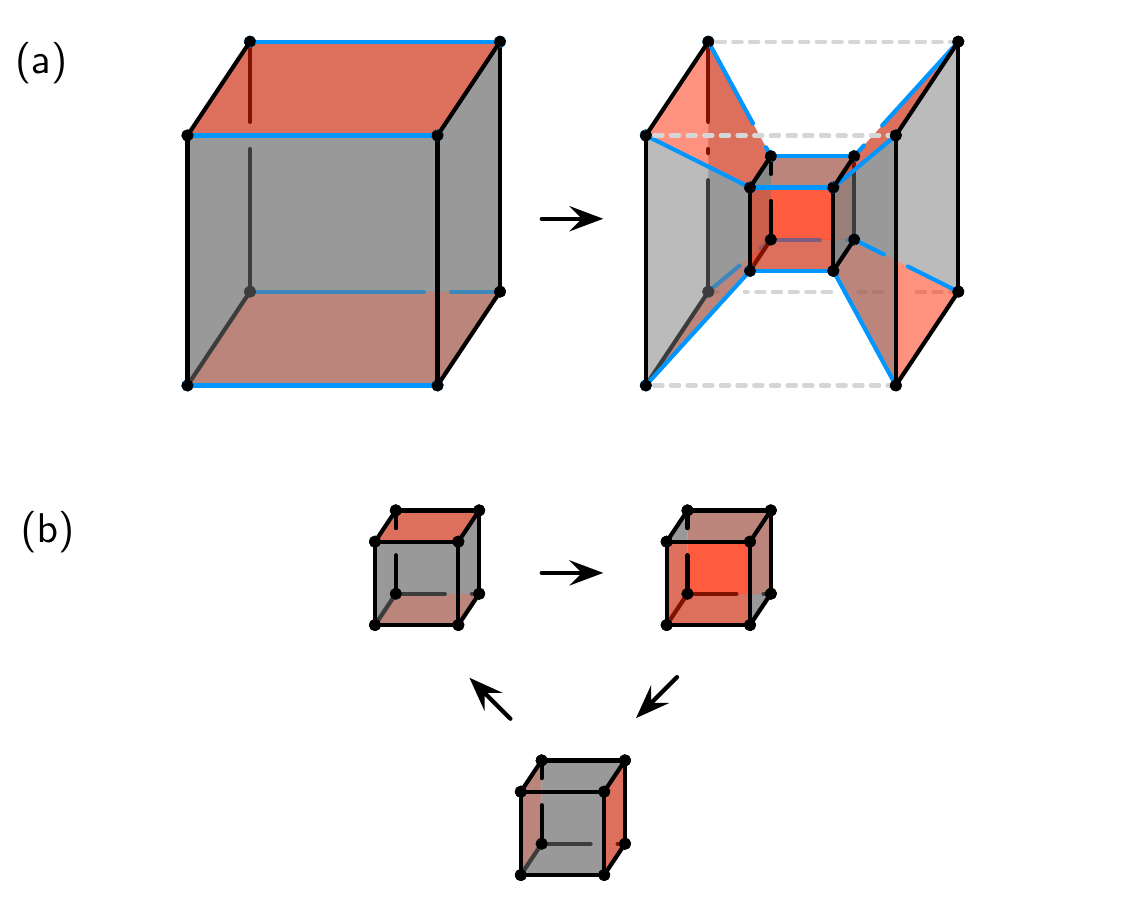}
\caption{(a) Triality mutation on a cubic brane brick in a brane brick model. (b) The triality move permutes fundamental and anti-fundamental chiral fields with Fermi fields, which has the effect of permuting the orientation of the cubic brane brick.}
\label{fig_02}
\end{figure}

The minimal gauge node in a brane brick model satisfying the constraints in \eref{es01a45} has, 
\beal{es01a46}
n_{k, \text{in}}^\chi =
n_{k, \text{out}}^\chi = n_k^\Lambda = 2 ~,~
\eea
and is referred to as the \textit{cubic node}.
The corresponding brane brick in the brane brick model is a cube with four chiral faces and two Fermi faces.
Triality acting on such a cubic brane brick is realized as a local rearrangement of the faces of the cube and its neighboring brane bricks, 
also referred to as a \textit{cubic move} \cite{Franco:2016nwv,Franco:2016qxh},
which is the brane brick analogue of the urban renewal move for brane tilings \cite{Franco:2005rj,Hanany:2005ss,Goncharov:2011hp,Kennaway:2007tq}, as illustrated in \fref{fig_02}.
\\

\paragraph{Quiver-Invariant Triality.}
Triality preserves the mesonic moduli space of a brane brick model
and hence identifies brane brick models corresponding to the same toric Calabi-Yau 4-fold \cite{Franco:2016nwv,Franco:2016qxh}.
A single triality transformation generally changes the quiver of the $2d$ $(0,2)$ theory.
This is because the mutation rules outlined above and illustrated in \fref{fig_02}
permute the field types between an incoming chiral, an outgoing chiral and a Fermi field
at the dualized gauge node. 
In general, the total numbers $n^\chi$ and $n^\Lambda$ also change in the dualized $2d$ $(0,2)$ theory.

In this work, we observe for the first time brane brick models corresponding to the same toric Calabi-Yau 4-fold that share
the same quiver and $E$-terms but have different $J$-terms.
We show that these models are related by a sequence of triality transformations at admissible gauge nodes satisfying \eref{es01a45}.
Although the intermediate quivers generally differ, the sequence returns the underlying quiver to its original form while changing the $J$- and $E$-terms.
We refer to this relation as \textit{quiver-invariant triality}.
\\

\subsection{Orbifold Reduction on Brane Tilings and $4d$ $\mathcal{N}=1$ Theories \label{sec023}}

\paragraph{Brane Tilings.}
The $4d$ $\mathcal{N}=1$ supersymmetric gauge theories realized as worldvolume theories of D3-branes probing toric Calabi-Yau 3-folds
are represented by bipartite graphs
on a 2-torus $T^2$, which are known as brane tilings \cite{Hanany:2005ve,Franco:2005rj,Franco:2005sm,Kennaway:2007tq}
or dimer models \cite{Hanany:2005ve,kenyon2003introductiondimermodel,kasteleyn1967graph,KENYON1997591}.
These bipartite graphs consist of faces
corresponding to gauge groups $U(N)_i$, 
edges corresponding to bifundamental chiral fields $\mathcal{X}_{ij}$,
and white and black nodes that correspond to positive and negative terms 
of the superpotential $W$, respectively.
The cyclic ordering of edges around a node 
specifies the ordering of chiral fields
in the corresponding term in $W$.
The bipartiteness of the graph ensures that 
every edge has a white and a black node as its endpoints and that 
every chiral field
appears exactly once in a positive term and once in a negative term of $W$.
Consequently, 
the $F$-term relations take the following binomial form, 
\beal{es01a50}
\frac{\partial W}{\partial \mathcal{X}_{ij}}
= M_{ji}^+ - M_{ji}^- = 0 ~,~
\eea
where $M^\pm_{ji}$ are ordered monomials in chiral fields.
For an abelian $4d$ $\mathcal{N}=1$ theory, these relations generate a binomial ideal whose zero locus is the master space \cite{Forcella:2008bb,Forcella:2008eh, Kho:2023dcm}.
Imposing the $D$-term constraints on its coherent component and quotienting by the gauge groups gives the
mesonic moduli space \cite{Forcella:2008bb,Franco:2006gc,Forcella:2008eh,Feng:2000mi}, which is the toric Calabi-Yau 3-fold probed by the D3-brane.

Brane tilings admit perfect matchings $p_a$ \cite{Hanany:2005ve,Franco:2005rj,Franco:2005sm,kenyon2003introductiondimermodel},
which are subsets of edges in the bipartite graph that cover every node exactly once. 
These perfect matchings $p_a$ correspond to GLSM fields \cite{Hanany:2005ve,Franco:2005rj,Franco:2005sm}
that parameterize the mesonic moduli space of the brane tiling.
\\

\paragraph{Dimensional Reduction.}
Ordinary dimensional reduction 
of a $4d$ $\mathcal{N}=1$ theory corresponding to a brane tiling 
gives
a $2d$ $(2,2)$ theory, which can be realized by a brane brick model.
The $4d$ vector $\mathcal{V}_i$
and chiral $\mathcal{X}_{ij}$ multiplets
become $2d$ $(2,2)$ vector and chiral multiplets.
Written in terms of $2d$ $(0,2)$ multiplets, 
the decomposition is as follows:
\begin{itemize}
\item a $4d$ vector multiplet $\mathcal{V}_i$ gives a $2d$ $(0,2)$ vector multiplet $V_i$ and an adjoint chiral multiplet $X_{ii}$,
\item a $4d$ chiral multiplet $\mathcal{X}_{ij}$ gives a $2d$ $(0,2)$ chiral multiplet $X_{ij}$ and a Fermi multiplet $\Lambda_{ij}$ transforming in the same representation.
\end{itemize}
For each Fermi $\Lambda_{ij}$, the $J$- and $E$-terms take the following form, 
\beal{es01a52}
J_{ji} = \frac{\partial W}{\partial \mathcal{X}_{ij}} ~,~
E_{ij} = X_{ii} X_{ij} - X_{ij} X_{jj} ~,~
\eea
where the derivative is the cyclic derivative of the parent $4d$ superpotential $W$ \cite{Franco:2015tna,Franco:2016fxm,Franco:2015tya}.
In the quiver, the $J$-term identifies paths along quiver edges from node $j$ to $i$, 
whereas the $E$-term identifies paths from node $i$ to $j$ as illustrated in \fref{fig_01}. 

On the coherent component of the abelian $2d$ theory, the equations $E_{ij}=(X_{ii}-X_{jj})X_{ij}=0$ imply that the adjoint chiral fields are equal on the connected quiver.
Accordingly, the mesonic moduli space of the $2d$ theory takes the following general form, 
\beal{es01a53}
\mathcal{M}^{mes}_{2d} = \text{CY}_3 \times \mathbb{C} ~,~
\eea
where the toric Calabi-Yau 3-fold $\text{CY}_3$ is the mesonic moduli space of the parent $4d$ $\mathcal{N}=1$ theory,
and $\mathbb{C}$ is parameterized by the common adjoint chiral field in $2d$. 

\begin{figure}[http!!]
\centering
\includegraphics[height=0.18\textheight]{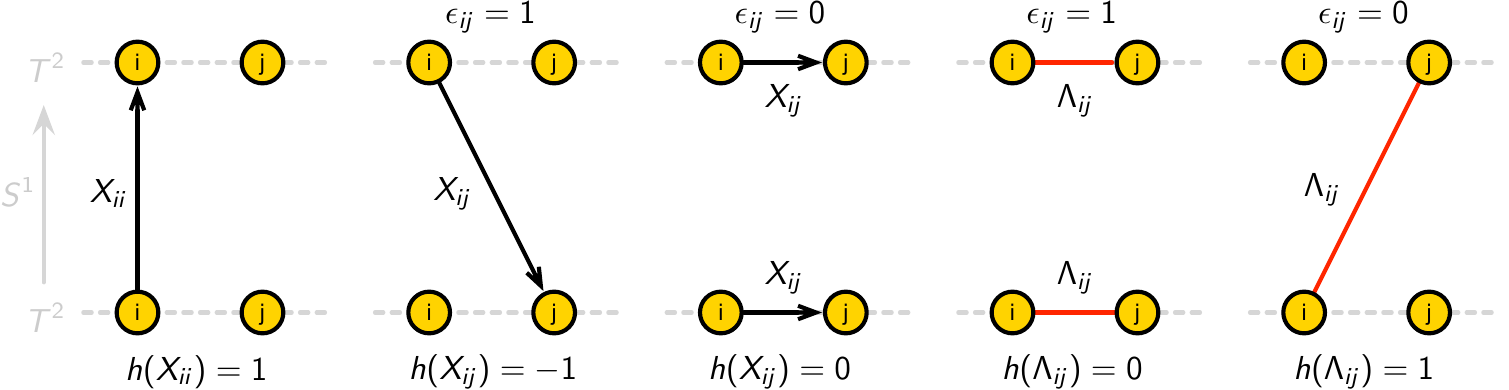}
\caption{
Values of height functions 
along the new $S^{1}$ direction for the $2d$ $(0,2)$ fields.
}
\label{fig_04}
\end{figure}

The above dimensional reduction from the $4d$ $\mathcal{N}=1$ theory
to the $2d$ $(2,2)$ theory expressed in terms of $(0,2)$
multiplets has a combinatorial realization as a lift of the periodic quiver from the 2-torus $T^2$ associated with the corresponding brane tiling, 
to a 3-torus $T^3$ associated with the brane brick model corresponding to the $2d$ theory.
Let us choose a perfect matching $p_0$ of the brane tiling
and define a $4d$ chiral field membership indicator of the perfect matching as follows, 
\beal{es01a55}
\epsilon_{ij}(p_0) = \left\{
\ba{cl}
1,~~ & \mathcal{X}_{ij} \in p_0
\\
0,~~ & \mathcal{X}_{ij} \notin p_0
\ea
\right.
~.~
\eea
We then add a new $S^1$ direction to the 2-torus $T^2$ of the brane tiling and assign the following height functions to the $2d$ fields,
\beal{es01a56}
h(X_{ii}) = 1 ~,~
h(X_{ij}) = - \epsilon_{ij} ~,~
h(\Lambda_{ij}) = 1 - \epsilon_{ij} ~,~
\eea
where \fref{fig_04} illustrates the above assignments.
Here, the heights of chiral fields $X_{ij}$ and $X_{ii}$
are measured from the tail to the head of the edge in the $2d$ periodic quiver on $T^3$, 
whereas for Fermi fields $\Lambda_{ij}$ we choose for bookkeeping the orientation from $i$ to $j$ inherited from $\mathcal{X}_{ij}$. 
Thus a $2d$ chiral field $X_{ij}$ coming from a $4d$ chiral field $\mathcal{X}_{ij}$
in $p_0$ is shifted in the $-1$ direction
and its $2d$ Fermi partner $\Lambda_{ij}$ remains along $T^2$.
Furthermore, a $2d$ chiral field $X_{ij}$ coming from a $4d$ chiral field $\mathcal{X}_{ij}$ that is not in $p_0$
remains along $T^2$ and its Fermi partner $\Lambda_{ij}$ is shifted in the $+1$ direction, as illustrated in \fref{fig_04}.

We further note that since every term in the $4d$ superpotential contains one chiral field in $p_0$,
each monomial of $J_{ji}$ has a displacement given by $\epsilon_{ij}- 1$
while each monomial of $E_{ij}$ has a displacement of $1-\epsilon_{ij}$.
Furthermore, the plaquettes containing
$\Lambda_{ij} J_{ji}$ and $\overline{\Lambda}_{ij} E_{ij}$
close in the resulting periodic quiver on $T^3$.
\\

\begin{figure}[http!!]
\centering
\includegraphics[height=0.46\textheight]{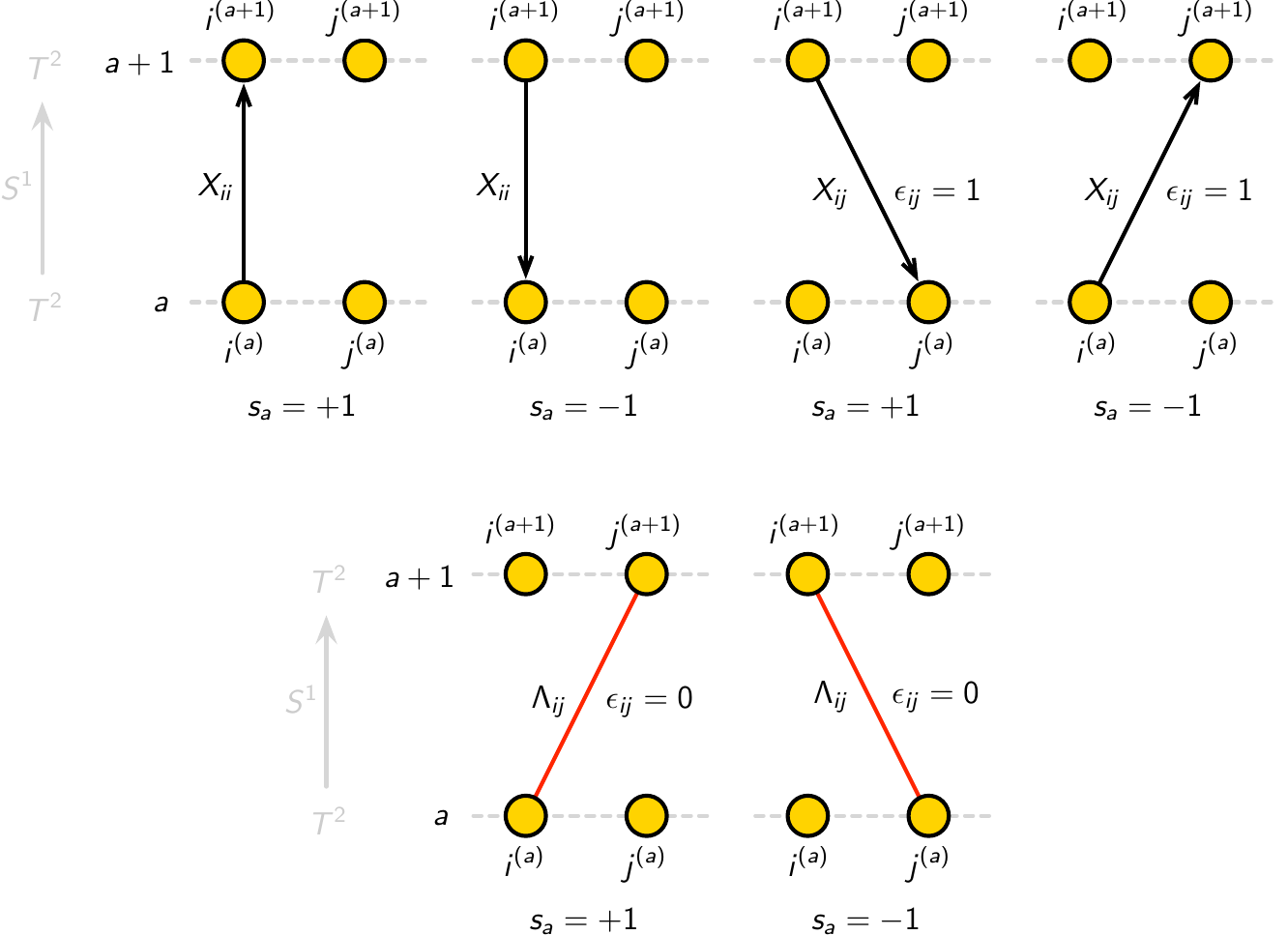}
\caption{
Shifts between layers along the new $S^1$ direction
of chiral and Fermi fields of the $2d$ $(0,2)$ theory during orbifold reduction with $s_a=\pm 1$.
}
\label{fig_05}
\end{figure}

\paragraph{Orbifold Reduction.}
Orbifold reduction generalizes the lift 
from $T^2$ to $T^3$ given by dimensional reduction
by introducing $k$ layered copies 
of the original periodic quiver on $T^2$
and allowing their relative orientations in $T^3$ to vary \cite{Franco:2016fxm}.
The input for orbifold reduction 
consists of a brane tiling on $T^2$
corresponding to a toric Calabi-Yau 3-fold,
a perfect matching $p_0$ in the brane tiling
and a sign vector given by,
\beal{es01a60}
s= (s_1, \dots, s_k)
\eea
where $s_a \in \{+1, -1 \}$.
With this input, 
we stack $k$ copies of the parent gauge nodes in the $S^1$ direction
and denote the copy of gauge node $i$ in layer $a$ by $i^{(a)}$.
We note here that layer indices $a$ are modulo $k$
such that the period along the new $S^1$ is $k$, as illustrated in \fref{fig_05}.

Within each $T^2$ layer, we retain a chiral field $X_{ij}$
for each parent field $\mathcal{X}_{ij} \notin p_0$ and a Fermi field $\Lambda_{ij}$ for each $\mathcal{X}_{ij} \in p_0$.
The fields connecting adjacent layers $a$ and $a+1$
are determined by the sign $s_a$.
The configuration for $s_a=-1$ is obtained from that
for $s_a=+1$ by reflection along the new $S^1$ direction.
Explicitly, these fields connect the gauge nodes in the new periodic quiver as follows, 
\beal{es01a61}
\ba{c|c|c}
\text{field} &
s_a = +1 &
s_a = -1
\\
\hline
X_{ii}
&
i^{(a)} \rightarrow i^{(a+1)}
&
i^{(a+1)} \rightarrow i^{(a)}
\\
X_{ij}, \quad \epsilon_{ij} = 1
&
i^{(a+1)} \rightarrow j^{(a)}
&
i^{(a)} \rightarrow j^{(a+1)}
\\
\Lambda_{ij}, \quad \epsilon_{ij} = 0
&
i^{(a)} \rightarrow j^{(a+1)}
&
i^{(a+1)} \rightarrow j^{(a)}
\ea
\eea
The resulting connections are illustrated in \fref{fig_05}.
Here, layer indices are understood modulo $k$.
As in \eref{es01a56}, we assign each Fermi field
$\Lambda_{ij}$ the orientation $i \to j$ inherited
from the corresponding parent chiral field $\mathcal{X}_{ij}$.
For $k>1$, the adjoint chiral fields $X_{ii}$
of the dimensionally reduced theory become bifundamental
fields connecting copies of the same parent gauge node
in adjacent layers.

The $J$- and $E$-terms are obtained from the plaquettes lifted along the new $S^1$ direction.
Each Fermi field is associated with two $J$-term plaquettes and two $E$-term plaquettes that yield
binomial relations satisfying the vanishing trace condition in \eref{es01a05}.
The periodic quiver on the resulting 3-torus $T^3$, which includes these plaquettes, determines the corresponding brane brick model \cite{Franco:2016fxm,Franco:2015tya,Franco:2015tna}
representing the associated $2d$ $(0,2)$ theory.

\begin{figure}[http!!]
\centering
\includegraphics[height=0.32\textheight]{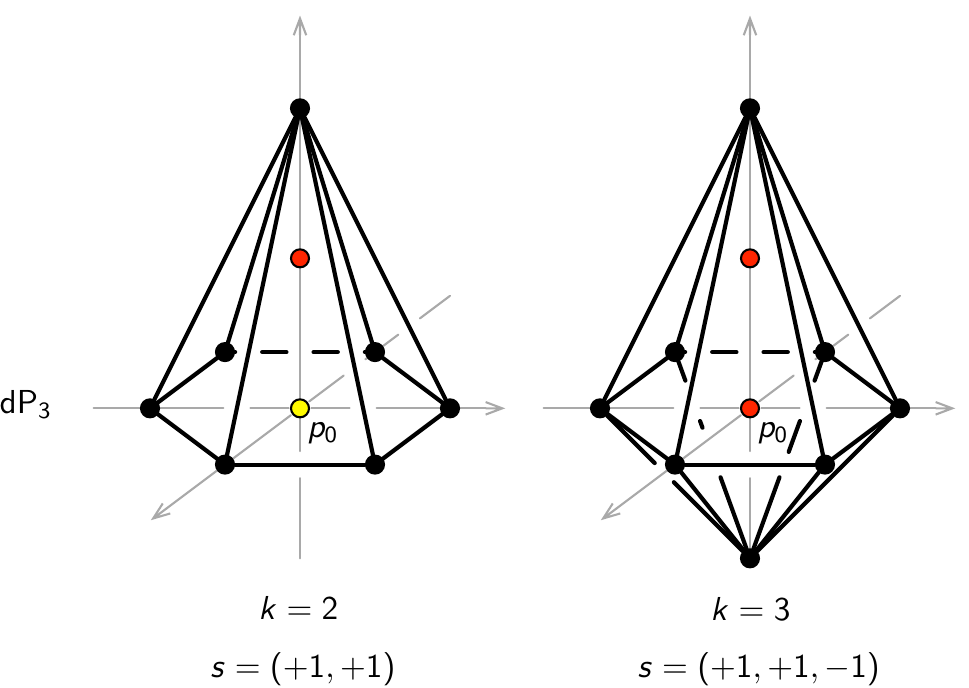}
\caption{
Effect of orbifold reduction on the toric diagram of the parent
Calabi-Yau 3-fold, here the cone over $\text{dP}_3$.
The point associated with the chosen perfect matching $p_0$ is extended into a vertical segment of lattice length
$k$.
}
\label{fig_06}
\end{figure}

Let $G_{4d}$ and $n^\chi_{4d}$ denote the numbers of gauge nodes and chiral multiplets in the parent $4d$ $\mathcal{N}=1$ theory.
Before integrating out any massive chiral-Fermi pairs, the numbers of gauge nodes, chiral fields and Fermi fields
in the orbifold-reduced theory are as follows,
\beal{es01a65}
G_{2d} = k G_{4d} ~,~
n^\chi_{2d} = k (n^\chi_{4d} + G_{4d}) ~,~
n^\Lambda_{2d} = k n^\chi_{4d} ~,~
\eea
where $k$ is the number of layers in the orbifold reduction.
We note that overall, the brane brick model satisfies,
\beal{es01a66}
n^\chi_{2d} - n^\Lambda_{2d} = G_{2d} ~.~
\eea
The local non-abelian anomaly cancellation condition must
also hold at every gauge node $i^{(a)}$, giving us the constraint,
\beal{es01a67}
n^\chi_{i^{(a)}} - n^\Lambda_{i^{(a)}} = 2 ~.~
\eea
This condition must be checked for the chosen perfect matching $p_0$ and sign vector $s$, since arbitrary choices need not
yield a consistent brane brick model \cite{Franco:2016fxm,Franco:2015tna}.

For a given choice of perfect matching $p_0$ and sign vector $s$,
orbifold reduction lifts the $2$-dimensional toric diagram of the parent toric Calabi-Yau 3-fold to a $3$-dimensional convex lattice polytope. 
This polytope represents the toric diagram of the toric Calabi-Yau 4-fold associated with the resulting brane brick model.
\fref{fig_06} illustrates how orbifold reduction lifts the toric diagram
of the parent Calabi-Yau 3-fold to that of the resulting Calabi-Yau 4-fold.
\\

\section{Quiver-Invariant Triality \label{sec03}}

\begin{figure}[htt!!!]
\centering
\begin{subfigure}[c]{0.5\textwidth}
\centering
\includegraphics[width=\linewidth]{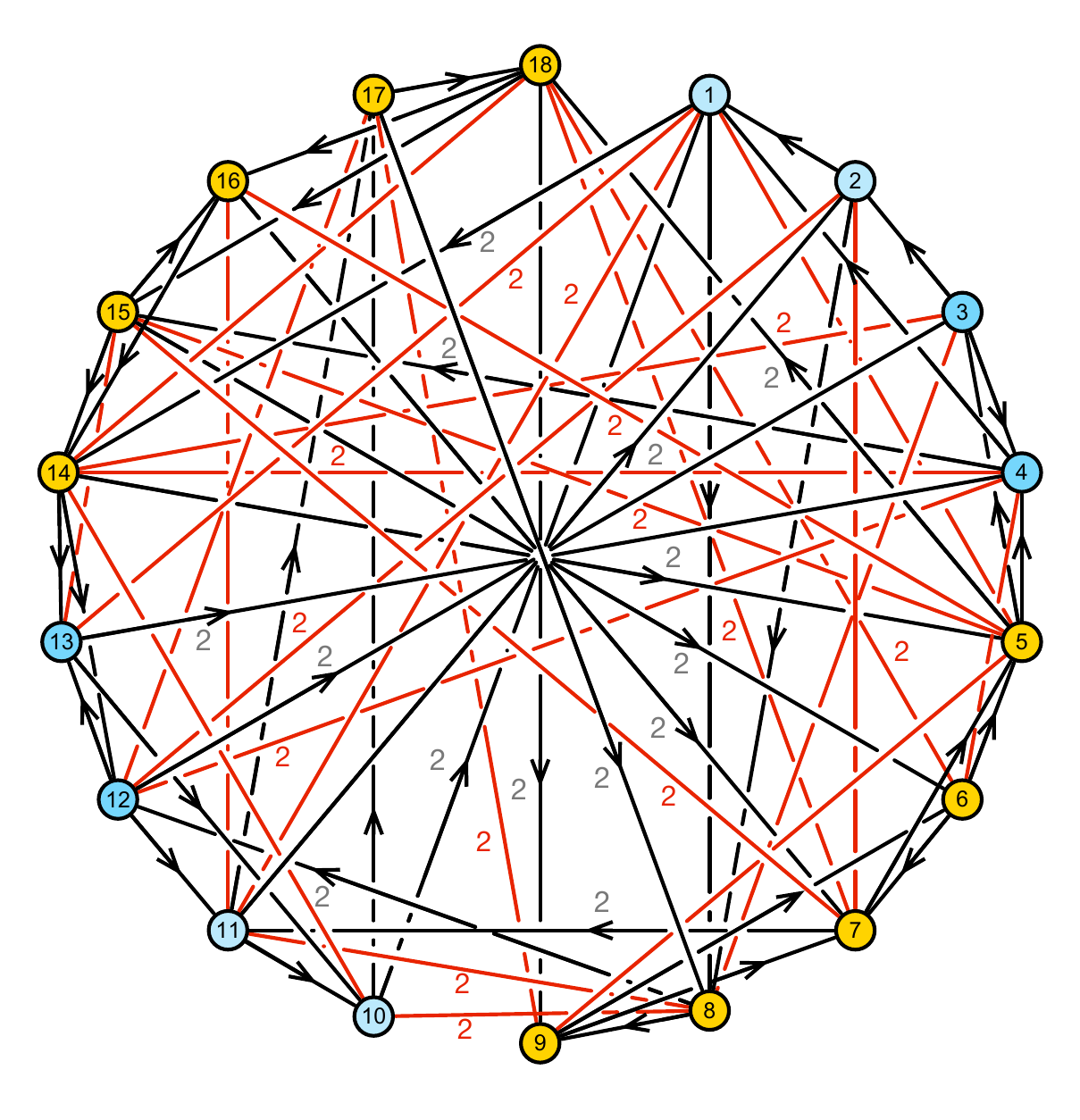}
\end{subfigure}
\hfill
\begin{subfigure}[c]{0.45\textwidth}
\centering
\includegraphics[width=\linewidth]{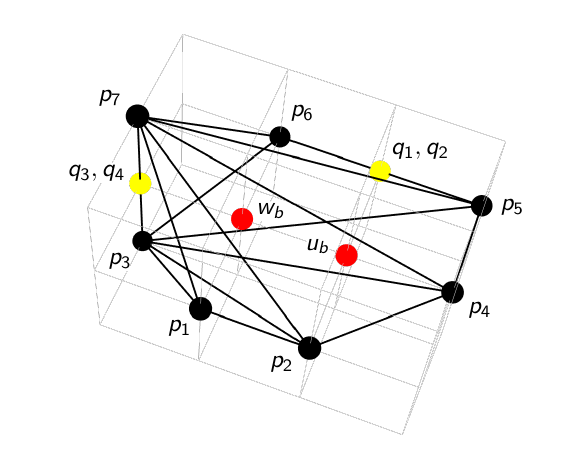}
\end{subfigure}
\caption{The quiver and toric diagram shared by Models A and B.}
\label{fig:modelAB}
\end{figure}

We present two brane brick models, denoted Models A and B, that share the same quiver and mesonic moduli space and have
identical $E$-terms but different $J$-terms.
Their common mesonic moduli space $\mathcal{M}^{mes}$ is a toric Calabi-Yau 4-fold.
We show in section \sref{sec:triality-chain} that the two models are related by a sequence of triality mutations.
Their common quiver and toric diagram are shown in \fref{fig:modelAB}.
\\

\begin{table}[htt!!!]
\centering
\begin{tabular}{|c|c|c|c|c|c|}
\hline
\; & $SU(2)_x$ & $U(1)_{f_1}$ & $U(1)_{f_2}$ & $U(1)_R$ & fugacity \\
\hline \hline
$p_1$ & $0$  & $+1$  	& $0$ 	& $R_1$ & $t_1 = f_1 t$ \\
$p_2$ & $0$ &  $+1$ 	& $0$ 	& $R_2$ & $t_2 = f_1 t$ \\
$p_3$ & $+1$   & $0$   & $0$ 	& $R_3$ & $t_{3} = x t$ \\
$p_4$ &  $0$   & $-2$    & $+2$ 	& $R_4$ & $t_4 = f_1^{-2} f_2^{2} t$ \\
$p_5$ &  $0$  &  $0$   & $-1$ 	& $R_5$ & $t_5 = f_2^{-1} t$ \\
$p_6$ &  $0$   & $0$    & $-1$ 	& $R_6$ & $t_6 = f_2^{-1} t$ \\
$p_7$ & $-1$   & $0$   & $0$  & $R_7$ & $t_7= x^{-1} t$ \\
\hline
\end{tabular}
\caption{
Global symmetry charges on the extremal GLSM fields
for Models A and B. 
\label{tab_01}
}
\end{table}

\subsection{Model A} \label{sec_modelA}

\begin{figure}[H]
\centering
\includegraphics[height=0.33\textheight]{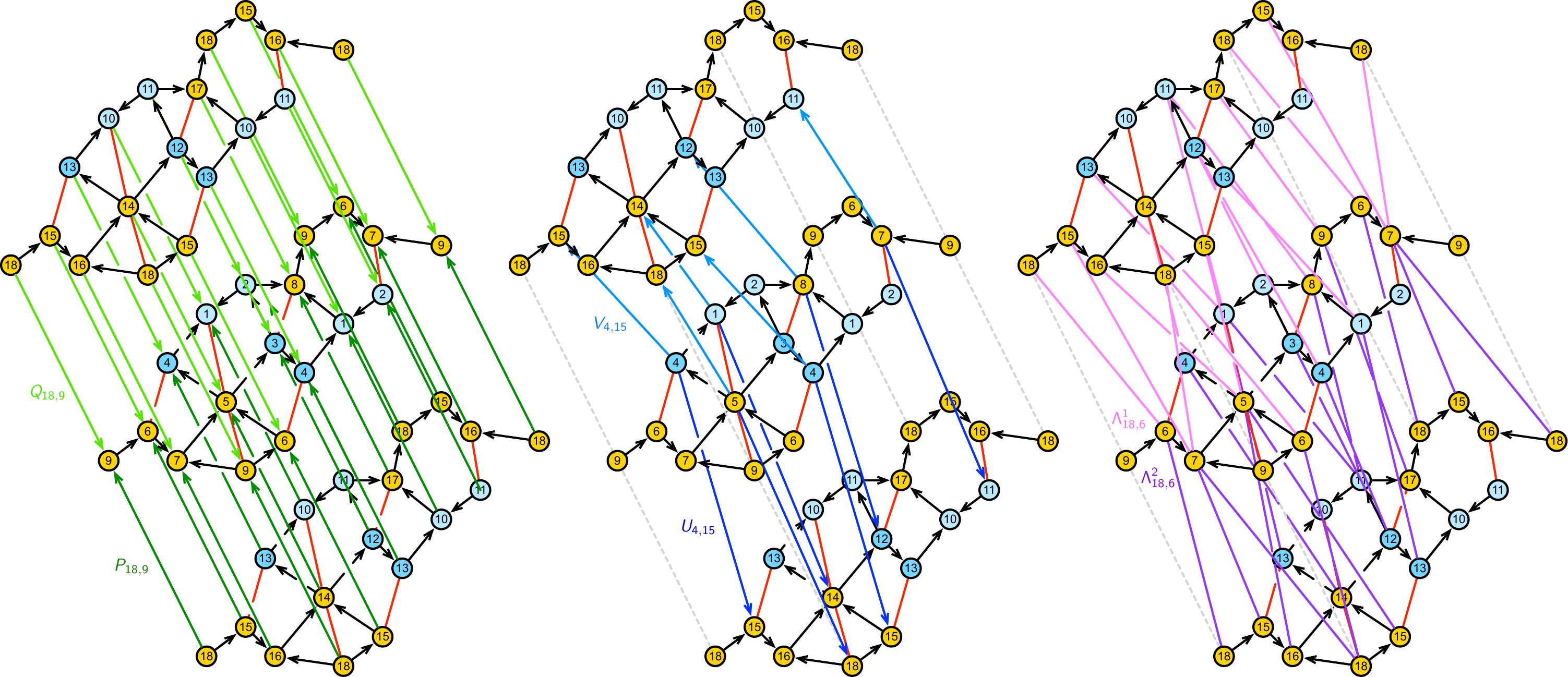}
\caption{The periodic quiver of Model A.}
\label{fig:periodic_quiver_A}
\end{figure}

The quiver for Model A is shown in \fref{fig:modelAB}. 
The periodic quiver, which encodes the $J$- and $E$-terms, 
is shown in \fref{fig:periodic_quiver_A}. 
The $J$- and $E$-terms of Model A are given by,
\beal{es02a01}
\resizebox{0.9\textwidth}{!}{$
\begin{array}{rrclcrcl}
& & J & & & & E &  \\
\Lambda_{15}: &
X_{54} \cdot X_{41} &-& X_{53} \cdot X_{32} \cdot X_{21}
& ~~ &
V_{1,14} \cdot Q_{14,5} &-& U_{1,14} \cdot P_{14,5}
\\
\Lambda_{83}: &
X_{32} \cdot X_{28} &-& X_{34} \cdot X_{41} \cdot X_{18}
& ~~ &
V_{8,12} \cdot Q_{12,3} &-& U_{8,12} \cdot P_{12,3}
\\
\Lambda_{46}: &
X_{65} \cdot X_{53} \cdot X_{34} &-& X_{67} \cdot X_{75} \cdot X_{54}
& ~~ &
V_{4,15} \cdot Q_{15,6} &-& U_{4,15} \cdot P_{15,6}
\\
\Lambda_{59}: &
X_{97} \cdot X_{75} &-& X_{96} \cdot X_{65}
& ~~ &
V_{5,18} \cdot Q_{18,9} &-& U_{5,18} \cdot P_{18,9}
\\
\Lambda_{72}: &
X_{21} \cdot X_{18} \cdot X_{89} \cdot X_{96} \cdot X_{67} &-& X_{28} \cdot X_{89} \cdot X_{97}
& ~~ &
V_{7,11} \cdot Q_{11,2} &-& U_{7,11} \cdot P_{11,2}
\\
\Lambda_{10,14}: &
X_{14,13} \cdot X_{13,10} &-& X_{14,12} \cdot X_{12,11} \cdot X_{11,10}
& ~~ &
P_{10,1} \cdot U_{1,14} &-& Q_{10,1} \cdot V_{1,14}
\\
\Lambda_{17,12}: &
X_{12,11} \cdot X_{11,17} &-& X_{12,13} \cdot X_{13,10} \cdot X_{10,17}
& ~~ &
P_{17,8} \cdot U_{8,12} &-& Q_{17,8} \cdot V_{8,12}
\\
\Lambda_{13,15}: &
X_{15,14} \cdot X_{14,12} \cdot X_{12,13} &-& X_{15,16} \cdot X_{16,14} \cdot X_{14,13}
& ~~ &
P_{13,4} \cdot U_{4,15} &-& Q_{13,4} \cdot V_{4,15}
\\
\Lambda_{14,18}: &
X_{18,16} \cdot X_{16,14} &-& X_{18,15} \cdot X_{15,14}
& ~~ &
P_{14,5} \cdot U_{5,18} &-& Q_{14,5} \cdot V_{5,18}
\\
\Lambda_{16,11}: &
X_{11,10} \cdot X_{10,17} \cdot X_{17,18} \cdot X_{18,15} \cdot X_{15,16} &-& X_{11,17} \cdot X_{17,18} \cdot X_{18,16}
& ~~ &
P_{16,7} \cdot U_{7,11} &-& Q_{16,7} \cdot V_{7,11}
\\
\Lambda^{1}_{10,8}: &
X_{89} \cdot X_{96} \cdot X_{67} \cdot V_{7,11} \cdot X_{11,10} &-& V_{8,12} \cdot X_{12,13} \cdot X_{13,10}
& ~~ &
X_{10,17} \cdot Q_{17,8} &-& Q_{10,1} \cdot X_{18}
\\
\Lambda^{1}_{11,1}: &
X_{18} \cdot X_{89} \cdot X_{96} \cdot X_{67} \cdot V_{7,11} &-& V_{1,14} \cdot X_{14,12} \cdot X_{12,11}
& ~~ &
X_{11,10} \cdot Q_{10,1} &-& Q_{11,2} \cdot X_{21}
\\
\Lambda^{1}_{11,8}: &
V_{8,12} \cdot X_{12,11} &-& X_{89} \cdot X_{97} \cdot V_{7,11}
& ~~ &
X_{11,17} \cdot Q_{17,8} &-& Q_{11,2} \cdot X_{28}
\\
\Lambda^{1}_{12,2}: &
X_{28} \cdot V_{8,12} &-& X_{21} \cdot V_{1,14} \cdot X_{14,12}
& ~~ &
X_{12,11} \cdot Q_{11,2} &-& Q_{12,3} \cdot X_{32}
\\
\Lambda^{1}_{12,4}: &
V_{4,15} \cdot X_{15,14} \cdot X_{14,12} &-& X_{41} \cdot X_{18} \cdot V_{8,12}
& ~~ &
X_{12,13} \cdot Q_{13,4} &-& Q_{12,3} \cdot X_{34}
\\
\Lambda^{1}_{13,1}: &
V_{1,14} \cdot X_{14,13} &-& X_{18} \cdot V_{8,12} \cdot X_{12,13}
& ~~ &
X_{13,10} \cdot Q_{10,1} &-& Q_{13,4} \cdot X_{41}
\\
\Lambda^{1}_{14,3}: &
X_{34} \cdot V_{4,15} \cdot X_{15,14} &-& X_{32} \cdot X_{21} \cdot V_{1,14}
& ~~ &
X_{14,12} \cdot Q_{12,3} &-& Q_{14,5} \cdot X_{53}
\\
\Lambda^{1}_{14,4}: &
X_{41} \cdot V_{1,14} &-& V_{4,15} \cdot X_{15,16} \cdot X_{16,14}
& ~~ &
X_{14,13} \cdot Q_{13,4} &-& Q_{14,5} \cdot X_{54}
\\
\Lambda^{1}_{15,5}: &
X_{53} \cdot X_{34} \cdot V_{4,15} &-& V_{5,18} \cdot X_{18,15}
& ~~ &
X_{15,14} \cdot Q_{14,5} &-& Q_{15,6} \cdot X_{65}
\\
\Lambda^{1}_{15,7}: &
V_{7,11} \cdot X_{11,10} \cdot X_{10,17} \cdot X_{17,18} \cdot X_{18,15} &-& X_{75} \cdot X_{54} \cdot V_{4,15}
& ~~ &
X_{15,16} \cdot Q_{16,7} &-& Q_{15,6} \cdot X_{67}
\\
\Lambda^{1}_{16,5}: &
V_{5,18} \cdot X_{18,16} &-& X_{54} \cdot V_{4,15} \cdot X_{15,16}
& ~~ &
X_{16,14} \cdot Q_{14,5} &-& Q_{16,7} \cdot X_{75}
\\
\Lambda^{1}_{17,9}: &
X_{96} \cdot X_{67} \cdot V_{7,11} \cdot X_{11,10} \cdot X_{10,17} &-& X_{97} \cdot V_{7,11} \cdot X_{11,17}
& ~~ &
X_{17,18} \cdot Q_{18,9} &-& Q_{17,8} \cdot X_{89}
\\
\Lambda^{1}_{18,6}: &
X_{67} \cdot V_{7,11} \cdot X_{11,10} \cdot X_{10,17} \cdot X_{17,18} &-& X_{65} \cdot V_{5,18}
& ~~ &
X_{18,15} \cdot Q_{15,6} &-& Q_{18,9} \cdot X_{96}
\\
\Lambda^{1}_{18,7}: &
X_{75} \cdot V_{5,18} &-& V_{7,11} \cdot X_{11,17} \cdot X_{17,18}
& ~~ &
X_{18,16} \cdot Q_{16,7} &-& Q_{18,9} \cdot X_{97}
\\
\Lambda^{2}_{10,8}: &
X_{89} \cdot X_{96} \cdot X_{67} \cdot U_{7,11} \cdot X_{11,10} &-& U_{8,12} \cdot X_{12,13} \cdot X_{13,10}
& ~~ &
P_{10,1} \cdot X_{18} &-& X_{10,17} \cdot P_{17,8}
\\
\Lambda^{2}_{11,1}: &
X_{18} \cdot X_{89} \cdot X_{96} \cdot X_{67} \cdot U_{7,11} &-& U_{1,14} \cdot X_{14,12} \cdot X_{12,11}
& ~~ &
P_{11,2} \cdot X_{21} &-& X_{11,10} \cdot P_{10,1}
\\
\Lambda^{2}_{11,8}: &
U_{8,12} \cdot X_{12,11} &-& X_{89} \cdot X_{97} \cdot U_{7,11}
& ~~ &
P_{11,2} \cdot X_{28} &-& X_{11,17} \cdot P_{17,8}
\\
\Lambda^{2}_{12,2}: &
X_{28} \cdot U_{8,12} &-& X_{21} \cdot U_{1,14} \cdot X_{14,12}
& ~~ &
P_{12,3} \cdot X_{32} &-& X_{12,11} \cdot P_{11,2}
\\
\Lambda^{2}_{12,4}: &
U_{4,15} \cdot X_{15,14} \cdot X_{14,12} &-& X_{41} \cdot X_{18} \cdot U_{8,12}
& ~~ &
P_{12,3} \cdot X_{34} &-& X_{12,13} \cdot P_{13,4}
\\
\Lambda^{2}_{13,1}: &
U_{1,14} \cdot X_{14,13} &-& X_{18} \cdot U_{8,12} \cdot X_{12,13}
& ~~ &
P_{13,4} \cdot X_{41} &-& X_{13,10} \cdot P_{10,1}
\\
\Lambda^{2}_{14,3}: &
X_{34} \cdot U_{4,15} \cdot X_{15,14} &-& X_{32} \cdot X_{21} \cdot U_{1,14}
& ~~ &
P_{14,5} \cdot X_{53} &-& X_{14,12} \cdot P_{12,3}
\\
\Lambda^{2}_{14,4}: &
X_{41} \cdot U_{1,14} &-& U_{4,15} \cdot X_{15,16} \cdot X_{16,14}
& ~~ &
P_{14,5} \cdot X_{54} &-& X_{14,13} \cdot P_{13,4}
\\
\Lambda^{2}_{15,5}: &
X_{53} \cdot X_{34} \cdot U_{4,15} &-& U_{5,18} \cdot X_{18,15}
& ~~ &
P_{15,6} \cdot X_{65} &-& X_{15,14} \cdot P_{14,5}
\\
\Lambda^{2}_{15,7}: &
U_{7,11} \cdot X_{11,10} \cdot X_{10,17} \cdot X_{17,18} \cdot X_{18,15} &-& X_{75} \cdot X_{54} \cdot U_{4,15}
& ~~ &
P_{15,6} \cdot X_{67} &-& X_{15,16} \cdot P_{16,7}
\\
\Lambda^{2}_{16,5}: &
U_{5,18} \cdot X_{18,16} &-& X_{54} \cdot U_{4,15} \cdot X_{15,16}
& ~~ &
P_{16,7} \cdot X_{75} &-& X_{16,14} \cdot P_{14,5}
\\
\Lambda^{2}_{17,9}: &
X_{96} \cdot X_{67} \cdot U_{7,11} \cdot X_{11,10} \cdot X_{10,17} &-& X_{97} \cdot U_{7,11} \cdot X_{11,17}
& ~~ &
P_{17,8} \cdot X_{89} &-& X_{17,18} \cdot P_{18,9}
\\
\Lambda^{2}_{18,6}: &
X_{67} \cdot U_{7,11} \cdot X_{11,10} \cdot X_{10,17} \cdot X_{17,18} &-& X_{65} \cdot U_{5,18}
& ~~ &
P_{18,9} \cdot X_{96} &-& X_{18,15} \cdot P_{15,6}
\\
\Lambda^{2}_{18,7}: &
X_{75} \cdot U_{5,18} &-& U_{7,11} \cdot X_{11,17} \cdot X_{17,18}
& ~~ &
P_{18,9} \cdot X_{97} &-& X_{18,16} \cdot P_{16,7}
\\
\end{array}
$}
~.~
\nn\\
\eea
The forward algorithm allows us to obtain the $P$-matrix, which is given by, 
\beal{es02a02}
P^A = 
\resizebox{0.55\textwidth}{!}{$
\left(
\begin{array}{c|ccccccc|cccc|ccc|ccc|ccc}
& p_1 & p_2 & p_3 & p_6 & p_4 & p_5 & p_7 & q_1 & q_2 & q_3 & q_4 & u_1 & \cdots & u_{42} & w_1 & \cdots & w_{65} & o^{(1)}_1 & \cdots & o^{(19)}_{83} \\
\hline
P_{10,1} & 0 & 0 & 1 & 0 & 0 & 0 & 0 & 0 & 0 & 0 & 1 & 0 & \cdots & 0 & 0 & \cdots & 0 & 0 & \cdots & 1 \\
 P_{11,2} & 0 & 0 & 1 & 0 & 0 & 0 & 0 & 0 & 0 & 0 & 1 & 0 & \cdots & 0 & 0 & \cdots & 0 & 0 & \cdots & 1 \\
 P_{12,3} & 0 & 0 & 1 & 0 & 0 & 0 & 0 & 0 & 0 & 0 & 1 & 0 & \cdots & 0 & 0 & \cdots & 0 & 0 & \cdots & 0 \\
 P_{13,4} & 0 & 0 & 1 & 0 & 0 & 0 & 0 & 0 & 0 & 0 & 1 & 0 & \cdots & 0 & 0 & \cdots & 0 & 0 & \cdots & 0 \\
 P_{14,5} & 0 & 0 & 1 & 0 & 0 & 0 & 0 & 0 & 0 & 0 & 1 & 0 & \cdots & 0 & 0 & \cdots & 1 & 0 & \cdots & 1 \\
 P_{15,6} & 0 & 0 & 1 & 0 & 0 & 0 & 0 & 0 & 0 & 0 & 1 & 0 & \cdots & 0 & 0 & \cdots & 1 & 0 & \cdots & 1 \\
 P_{16,7} & 0 & 0 & 1 & 0 & 0 & 0 & 0 & 0 & 0 & 0 & 1 & 0 & \cdots & 0 & 0 & \cdots & 1 & 0 & \cdots & 2 \\
 P_{17,8} & 0 & 0 & 1 & 0 & 0 & 0 & 0 & 0 & 0 & 0 & 1 & 1 & \cdots & 0 & 1 & \cdots & 0 & 1 & \cdots & 0 \\
 P_{18,9} & 0 & 0 & 1 & 0 & 0 & 0 & 0 & 0 & 0 & 0 & 1 & 0 & \cdots & 1 & 0 & \cdots & 1 & 0 & \cdots & 2 \\
 Q_{10,1} & 0 & 0 & 0 & 0 & 0 & 0 & 1 & 0 & 0 & 0 & 1 & 0 & \cdots & 0 & 0 & \cdots & 0 & 0 & \cdots & 1 \\
 Q_{11,2} & 0 & 0 & 0 & 0 & 0 & 0 & 1 & 0 & 0 & 0 & 1 & 0 & \cdots & 0 & 0 & \cdots & 0 & 0 & \cdots & 1 \\
 Q_{12,3} & 0 & 0 & 0 & 0 & 0 & 0 & 1 & 0 & 0 & 0 & 1 & 0 & \cdots & 0 & 0 & \cdots & 0 & 0 & \cdots & 0 \\
 Q_{13,4} & 0 & 0 & 0 & 0 & 0 & 0 & 1 & 0 & 0 & 0 & 1 & 0 & \cdots & 0 & 0 & \cdots & 0 & 0 & \cdots & 0 \\
 Q_{14,5} & 0 & 0 & 0 & 0 & 0 & 0 & 1 & 0 & 0 & 0 & 1 & 0 & \cdots & 0 & 0 & \cdots & 1 & 0 & \cdots & 1 \\
 Q_{15,6} & 0 & 0 & 0 & 0 & 0 & 0 & 1 & 0 & 0 & 0 & 1 & 0 & \cdots & 0 & 0 & \cdots & 1 & 0 & \cdots & 1 \\
 Q_{16,7} & 0 & 0 & 0 & 0 & 0 & 0 & 1 & 0 & 0 & 0 & 1 & 0 & \cdots & 0 & 0 & \cdots & 1 & 0 & \cdots & 2 \\
 Q_{17,8} & 0 & 0 & 0 & 0 & 0 & 0 & 1 & 0 & 0 & 0 & 1 & 1 & \cdots & 0 & 1 & \cdots & 0 & 1 & \cdots & 0 \\
 Q_{18,9} & 0 & 0 & 0 & 0 & 0 & 0 & 1 & 0 & 0 & 0 & 1 & 0 & \cdots & 1 & 0 & \cdots & 1 & 0 & \cdots & 2 \\
 U_{1,14} & 0 & 0 & 0 & 1 & 0 & 0 & 1 & 1 & 0 & 1 & 0 & 1 & \cdots & 0 & 1 & \cdots & 0 & 0 & \cdots & 0 \\
 U_{4,15} & 0 & 0 & 0 & 0 & 0 & 0 & 1 & 0 & 0 & 1 & 0 & 0 & \cdots & 0 & 1 & \cdots & 0 & 0 & \cdots & 0 \\
 U_{5,18} & 0 & 0 & 0 & 1 & 0 & 0 & 1 & 0 & 1 & 1 & 0 & 0 & \cdots & 0 & 1 & \cdots & 0 & 0 & \cdots & 2 \\
 U_{7,11} & 0 & 0 & 0 & 0 & 0 & 0 & 1 & 0 & 0 & 1 & 0 & 0 & \cdots & 0 & 0 & \cdots & 0 & 0 & \cdots & 0 \\
 U_{8,12} & 1 & 0 & 0 & 0 & 0 & 0 & 1 & 0 & 0 & 1 & 0 & 0 & \cdots & 1 & 0 & \cdots & 1 & 0 & \cdots & 0 \\
 V_{1,14} & 0 & 0 & 1 & 1 & 0 & 0 & 0 & 1 & 0 & 1 & 0 & 1 & \cdots & 0 & 1 & \cdots & 0 & 0 & \cdots & 0 \\
 V_{4,15} & 0 & 0 & 1 & 0 & 0 & 0 & 0 & 0 & 0 & 1 & 0 & 0 & \cdots & 0 & 1 & \cdots & 0 & 0 & \cdots & 0 \\
 V_{5,18} & 0 & 0 & 1 & 1 & 0 & 0 & 0 & 0 & 1 & 1 & 0 & 0 & \cdots & 0 & 1 & \cdots & 0 & 0 & \cdots & 2 \\
 V_{7,11} & 0 & 0 & 1 & 0 & 0 & 0 & 0 & 0 & 0 & 1 & 0 & 0 & \cdots & 0 & 0 & \cdots & 0 & 0 & \cdots & 0 \\
 V_{8,12} & 1 & 0 & 1 & 0 & 0 & 0 & 0 & 0 & 0 & 1 & 0 & 0 & \cdots & 1 & 0 & \cdots & 1 & 0 & \cdots & 0 \\
 X_{18} & 0 & 1 & 0 & 0 & 0 & 0 & 0 & 0 & 0 & 0 & 0 & 1 & \cdots & 0 & 1 & \cdots & 0 & 1 & \cdots & 0 \\
 X_{21} & 1 & 0 & 0 & 0 & 0 & 0 & 0 & 0 & 0 & 0 & 0 & 0 & \cdots & 0 & 0 & \cdots & 0 & 0 & \cdots & 0 \\
 X_{28} & 0 & 0 & 0 & 1 & 0 & 1 & 0 & 1 & 1 & 0 & 0 & 1 & \cdots & 0 & 1 & \cdots & 0 & 1 & \cdots & 1 \\
 X_{32} & 0 & 1 & 0 & 0 & 1 & 0 & 0 & 0 & 0 & 0 & 0 & 0 & \cdots & 0 & 0 & \cdots & 0 & 1 & \cdots & 3 \\
 X_{34} & 0 & 0 & 0 & 1 & 0 & 0 & 0 & 1 & 0 & 0 & 0 & 0 & \cdots & 0 & 0 & \cdots & 0 & 0 & \cdots & 3 \\
 X_{41} & 0 & 0 & 0 & 0 & 1 & 1 & 0 & 0 & 1 & 0 & 0 & 0 & \cdots & 0 & 0 & \cdots & 0 & 1 & \cdots & 1 \\
 X_{53} & 0 & 0 & 0 & 0 & 0 & 1 & 0 & 0 & 1 & 0 & 0 & 0 & \cdots & 1 & 0 & \cdots & 0 & 1 & \cdots & 0 \\
 X_{54} & 1 & 1 & 0 & 0 & 0 & 0 & 0 & 0 & 0 & 0 & 0 & 0 & \cdots & 1 & 0 & \cdots & 0 & 1 & \cdots & 2 \\
 X_{65} & 1 & 1 & 0 & 0 & 1 & 0 & 0 & 0 & 0 & 0 & 0 & 1 & \cdots & 0 & 0 & \cdots & 0 & 1 & \cdots & 0 \\
 X_{67} & 0 & 0 & 0 & 1 & 0 & 0 & 0 & 0 & 1 & 0 & 0 & 0 & \cdots & 0 & 0 & \cdots & 0 & 0 & \cdots & 1 \\
 X_{75} & 0 & 0 & 0 & 0 & 1 & 1 & 0 & 1 & 0 & 0 & 0 & 1 & \cdots & 0 & 0 & \cdots & 0 & 1 & \cdots & 0 \\
 X_{89} & 0 & 0 & 0 & 0 & 1 & 0 & 0 & 0 & 0 & 0 & 0 & 0 & \cdots & 1 & 0 & \cdots & 1 & 0 & \cdots & 2 \\
 X_{96} & 0 & 0 & 0 & 0 & 0 & 1 & 0 & 1 & 0 & 0 & 0 & 0 & \cdots & 0 & 0 & \cdots & 0 & 1 & \cdots & 0 \\
 X_{97} & 1 & 1 & 0 & 0 & 0 & 0 & 0 & 0 & 0 & 0 & 0 & 0 & \cdots & 0 & 0 & \cdots & 0 & 1 & \cdots & 0 \\
 X_{10,17} & 0 & 1 & 0 & 0 & 0 & 0 & 0 & 0 & 0 & 0 & 0 & 0 & \cdots & 0 & 0 & \cdots & 0 & 0 & \cdots & 1 \\
 X_{11,10} & 1 & 0 & 0 & 0 & 0 & 0 & 0 & 0 & 0 & 0 & 0 & 0 & \cdots & 0 & 0 & \cdots & 0 & 0 & \cdots & 0 \\
 X_{11,17} & 0 & 0 & 0 & 1 & 0 & 1 & 0 & 1 & 1 & 0 & 0 & 0 & \cdots & 0 & 0 & \cdots & 0 & 0 & \cdots & 2 \\
 X_{12,11} & 0 & 1 & 0 & 0 & 1 & 0 & 0 & 0 & 0 & 0 & 0 & 0 & \cdots & 0 & 0 & \cdots & 0 & 1 & \cdots & 2 \\
 X_{12,13} & 0 & 0 & 0 & 1 & 0 & 0 & 0 & 1 & 0 & 0 & 0 & 0 & \cdots & 0 & 0 & \cdots & 0 & 0 & \cdots & 3 \\
 X_{13,10} & 0 & 0 & 0 & 0 & 1 & 1 & 0 & 0 & 1 & 0 & 0 & 0 & \cdots & 0 & 0 & \cdots & 0 & 1 & \cdots & 0 \\
 X_{14,12} & 0 & 0 & 0 & 0 & 0 & 1 & 0 & 0 & 1 & 0 & 0 & 0 & \cdots & 1 & 0 & \cdots & 1 & 1 & \cdots & 1 \\
 X_{14,13} & 1 & 1 & 0 & 0 & 0 & 0 & 0 & 0 & 0 & 0 & 0 & 0 & \cdots & 1 & 0 & \cdots & 1 & 1 & \cdots & 3 \\
 X_{15,14} & 1 & 1 & 0 & 0 & 1 & 0 & 0 & 0 & 0 & 0 & 0 & 1 & \cdots & 0 & 0 & \cdots & 0 & 1 & \cdots & 0 \\
 X_{15,16} & 0 & 0 & 0 & 1 & 0 & 0 & 0 & 0 & 1 & 0 & 0 & 0 & \cdots & 0 & 0 & \cdots & 0 & 0 & \cdots & 0 \\
 X_{16,14} & 0 & 0 & 0 & 0 & 1 & 1 & 0 & 1 & 0 & 0 & 0 & 1 & \cdots & 0 & 0 & \cdots & 0 & 1 & \cdots & 1 \\
 X_{17,18} & 0 & 0 & 0 & 0 & 1 & 0 & 0 & 0 & 0 & 0 & 0 & 1 & \cdots & 0 & 1 & \cdots & 0 & 1 & \cdots & 0 \\
 X_{18,15} & 0 & 0 & 0 & 0 & 0 & 1 & 0 & 1 & 0 & 0 & 0 & 0 & \cdots & 1 & 0 & \cdots & 0 & 1 & \cdots & 1 \\
 X_{18,16} & 1 & 1 & 0 & 0 & 0 & 0 & 0 & 0 & 0 & 0 & 0 & 0 & \cdots & 1 & 0 & \cdots & 0 & 1 & \cdots & 0 \\
\end{array}
\right)
$}
~,~
\eea
where $p_a$ correspond to GLSM fields associated to extremal vertices of the toric diagram shown in \fref{fig:modelAB}. 
The columns denoted by $o$ correspond to extra GLSM fields \cite{Franco:2015tna,Franco:2015tya}. 
The toric diagram for Model A is obtained from the following coordinate matrix,
\beal{es02a03}
G_t^A=
\resizebox{0.8\textwidth}{!}{$
\left(
\begin{array}{ccccccc|cccc|ccc|ccc|ccc}
p_1 & p_2 & p_3  & p_4 & p_5 & p_6 & p_7 & q_1 & q_2 & q_3 & q_4 & u_1 & \cdots & u_{42} & w_1 & \cdots & w_{65} & o^{(1)}_1 & \cdots & o^{(19)}_{83} \\
\hline
 0 & 1 & -1 & 2 & 2 & 0 & -1 & 1 & 1 & -1 & -1 & 1 & \cdots & 1 & 0 & \cdots & 0 & 3 & \cdots & 2 \\
 -1 & -1 & 0 & 0 & 1 & 1 & 0 & 1 & 1 & 0 & 0 & 0 & \cdots & 0 & 0 & \cdots & 0 & 0 & \cdots & 1 \\
 0 & 0 & 1 & 0 & 0 & 0 & -1 & 0 & 0 & 0 & 0 & 0 &\cdots &  0 & 0 & \cdots & 0 & 0 & \cdots & 0 \\
 \hline
 1 & 1 & 1 & 1 & 1 & 1 & 1 & 1 & 1 & 1 & 1 & 1 & \cdots & 1 & 1 & \cdots & 1 & 2 & \cdots & 4 \\
\end{array}
\right)
$}
~.~
\eea

The enhanced global symmetry of Model A is given by, 
\beal{es02a04}
SU(2)_x \times U(1)_{f_1} \times U(1)_{f_2} \times U(1)_R
~.~
\eea
The Hilbert series of the mesonic moduli space $\mathcal{M}_A^{mes}$ is refined by the mesonic flavor fugacities and the auxiliary
grading fugacity $t$, as summarized in \tref{tab_01}.
The highest weight generating function of the mesonic moduli space $\mathcal{M}_A^{mes}$ 
then takes the following form,
\beal{es02a05}
h(\mu,f_1,f_2,t; \mathcal{M}^{mes}_A)
&& 
=\frac{P(\mu,f_1,f_2,t; \mathcal{M}^{mes}_A)}{(1-\mu^2 f^2_1 f^2_2 t^7)(1- \mu^2 f^{-2}_1 f^{-2}_2 t^7) (1-{f_1}^{-5} t^{10}) }
\nn\\
&&
\hspace{1cm}
\times \frac{1}{(1 - {f_1}^{-1} f_2^4 t^{10})(1- \mu^6 f_1 {f_2}^{-4} t^{11}) (1- \mu^6 {f_1}^3 {f_2}^{-2} t^{11})}
~,~
\nn\\
\eea
where $t$ is the fugacity counting degrees of the extremal GLSM fields $p_a$, 
and $\mu^m$ corresponds to the $SU(2)_x$ irreducible representation of the form $[m]_x$ with highest weight $(m)$.
The fugacities $f_1$ and $f_2$ are associated to $U(1)_{f_1}$ and $U(1)_{f_2}$ factors of the mesonic flavor symmetry, respectively.
The numerator of the generating function in \eref{es02a05} is given by, 
\beal{es02a06}
&&
P(\mu,f_1,f_2,t; \mathcal{M}^{mes}_A) = 
1 + \mu^{2} t^{7} + {f_1}^{-3} {f_2}^{2} t^{10} 
\nn\\
&&
\hspace{1cm}
- \mu^{2} {f_1}^{-5} t^{17} - \mu^{2} {f_1}^{-3} {f_2}^{2} t^{17} - \mu^{2} {f_1}^{-1} {f_2}^{4} t^{17} 
- \mu^{8} {f_1} {f_2}^{-4} t^{18} - \mu^{8} {f_1}^{3} {f_2}^{-2} t^{18}
\nn\\
&&
\hspace{1cm}
- 2 \mu^{6} t^{21} - \mu^{6} {f_1}^{-4} {f_2}^{-4} t^{21} - 2 \mu^{6} {f_1}^{-2} {f_2}^{-2} t^{21} - \mu^{6} {f_1}^{2} {f_2}^{2} t^{21}
\nn\\
&&
\hspace{1cm}
 + 2 \mu^{8} t^{28} + \mu^{8} {f_1}^{-4} {f_2}^{-4} t^{28} + 2 \mu^{8} {f_1}^{-2} {f_2}^{-2} t^{28} + \mu^{8} {f_1}^{2} {f_2}^{2} t^{28}
\nn\\
&&
 \hspace{1cm}
 + \mu^{6} {f_1}^{-5} t^{31} + \mu^{6} {f_1}^{-3} {f_2}^{2} t^{31}   + \mu^{12} {f_1}^{-1} {f_2}^{-6} t^{32}
  + \mu^{12} {f_1} {f_2}^{-4} t^{32} + \mu^{12} {f_1}^{3} {f_2}^{-2} t^{32}
 \nn\\
 &&
 \hspace{1cm}
 - \mu^{14} {f_1} {f_2}^{-4} t^{39}
 - \mu^{12} {f_1}^{-2} {f_2}^{-2} t^{42} - \mu^{14} {f_1}^{-2} {f_2}^{-2} t^{49}
~.~
\eea
The refined Hilbert series is obtained by replacing each monomial $\mu^m$ in the highest weight expansion of
\eref{es02a05} by the $SU(2)_x$ character $[m]_x=\sum_{j=0}^{m}x^{m-2j}$.
Setting $x=f_1=f_2=1$ then gives the unrefined Hilbert series of the following form,
\beal{es06a07}
g(t) = \frac{P(t,\mathcal{M}^{mes}_A) (1-t)^2 (1-t^2)}{(1-t^7)^2 (1-t^{10})^2 (1-t^{11})^3}
~,~
\eea
where the palindromic polynomial $P(t,\mathcal{M}^{mes}_A)$ is given by, 
\beal{eq:modelA-numerator}
&&
P(t,\mathcal{M}^{mes}_A)=
1 + 2 t + 4 t^{2} + 6 t^{3} + 9 t^{4} + 12 t^{5} + 16 t^{6} + 27 t^{7} + 39 t^{8} + 58 t^{9} 
\nn\\
&&
\hspace{1cm}
+ 79 t^{10} + 118 t^{11} + 159 t^{12} + 
218 t^{13} + 287 t^{14} + 374 t^{15} + 471 t^{16} + 581 t^{17} 
 \nn\\
&&
\hspace{1cm}
+ 688 
t^{18} + 808 t^{19} + 925 t^{20} + 1032 t^{21} + 1136 t^{22} 
+ 1230 t^{23} + 1313 t^{24} + 1370 t^{25}
\nn\\
&&
\hspace{1cm}
 + 1416 t^{26} + 1436 t^{27} 
+ 1452 
t^{28} + 1436 t^{29} + 1416 t^{30} 
\nn\\
&& 
\hspace{1cm}
+ 1370 t^{31} + 1313 t^{32} 
+ 1230 
t^{33} + 1136 t^{34} + 1032 t^{35} + 925 t^{36} + 808 t^{37} 
+ 688 t^{38} 
\nn\\
&&
\hspace{1cm}
+ 581 t^{39}  + 471 t^{40} + 374 t^{41} + 287 t^{42} 
+ 218 
t^{43} + 159 t^{44} + 118 t^{45} + 79 t^{46} 
\nn\\
&&
\hspace{1cm}
+ 58 t^{47} + 39 t^{48} 
+ 27 t^{49} 
+ 16 t^{50} + 12 t^{51} + 9 t^{52} + 6 t^{53} + 4 t^{54} 
+ 
2 t^{55} + t^{56}
~.~
\eea

The plethystic logarithm of the refined Hilbert series
has the following expansion, 
\beal{es02a07}
&&\text{PL}[g(x,f_1,f_2,t;\mathcal{M}_A^{mes})]= ( [2]_x f_1^{2} f_2^{2} + [2]_x + [2]_x f_1^{-2} f_2^{-2}) t^7
\nn\\\
&&
\hspace{1cm}
+( f_1^{-1} f_2^4 +f_1^{-3} f_2^2 + f_1^{-5}  ) t^{10} 
+  ( [6]_x f_1^3 f_2^{-2} + [6]_x f_1 f_2^{-4} ) t^{11}
\nn\\
&&
\hspace{1cm}
-( ([4]_x + [2]_x f_1^{2} f_2^{2} + [2]_x + [2]_x f_1^{-2} f_2^{-2} 
\nn\\
&&
\hspace{1cm}
+ f_1^{4} f_2^{4} + f_1^{2} f_2^{2} +2 + f_1^{-2} f_2^{-2} + f_1^{-4} f_2^{-4}
 ) t^{14}
\nn\\
&&
\hspace{1cm}
+ ([2]_x f_1^{-1} f_2^4  + 2 [2]_x f_1^{-3} f_2^2 + [2]_x f_1^{-5} ) t^{17}
\nn\\
&&
\hspace{1cm}
+( [8]_x f_1^3 f_2^{-2} + [8]_x f_1 f_2^{-4} + [6]_x f_1^5  + 2 [6]_x f_1^3 f_2^{-2} + 2 [6]_x f_1 f_2^{-4}
\nn\\
&&
\hspace{1cm}
  + [6]_x f_1^{-1} f_2^{-6} + [4]_x f_1^5 + 2[4]_x f_1^3 f_2^{-2} + 2[4]_x f_1 f_2^{-4} + [4]_x f_1^{-1} f_2^{-6} ) t^{18} 
 \nn\\
&&
\hspace{1cm}
+ f_1^{-6} f_2^4 t^{20}) + \dots
~,~
\eea
where the
leading positive terms correspond to the $26$ generators of the mesonic moduli space $\mathcal{M}_A^{mes}$. 
Here, we note that $\mathcal{M}_A^{mes}$ is not a complete intersection
as indicated by the infinite expansion of the plethystic logarithm.
\\

\begin{table}[H]
\centering
\newcommand{\chunkline}{\cline{1-5}\cline{7-11}}
\resizebox{0.8\textwidth}{!}{
\begin{tabular}{|c|c|c|c|l| p{1em} |c|c|c|c|l|}
\chunkline
\; & $SU(2)_x$ & $U(1)_{f_1}$ & $U(1)_{f_2}$ & $U(1)_R$ &
   & \; & $SU(2)_x$ & $U(1)_{f_1}$ & $U(1)_{f_2}$ & $U(1)_R$ \\
\chunkline \noalign{\vskip\doublerulesep} \chunkline
$P_{10,1}$ & $+1$ & $0$ & $0$ & $R_{3}$ &
	& $Q_{10,1}$ & $-1$ & $0$ & $0$ & $R_{7}$ \\
\chunkline
$P_{11,2}$ & $+1$ & $0$ & $0$ & $R_{3}$  &
	& $Q_{11,2}$ & $-1$ & $0$ & $0$ & $R_{7}$  \\
\chunkline
$P_{12,3}$ & $+1$ & $0$ & $0$ & $R_{3}$ &
	& $Q_{12,3}$ & $-1$ & $0$ & $0$ & $R_{7}$  \\
\chunkline
$P_{13,4}$ & $+1$ & $0$ & $0$ & $R_{3}$ &
	& $Q_{13,4}$ & $-1$ & $0$ & $0$ & $R_{7}$  \\
\chunkline
$P_{14,5}$ & $+1$ & $0$ & $0$ & $R_{3}$ &
	&  $Q_{14,5}$ & $-1$ & $0$ & $0$ & $R_{7}$ \\
\chunkline
$P_{15,6}$ & $+1$ & $0$ & $0$ & $R_{3}$ &
	& $Q_{15,6}$ & $-1$ & $0$ & $0$ & $R_{7}$  \\
\chunkline
$P_{16,7}$ & $+1$ & $0$ & $0$ & $R_{3}$ &
	& $Q_{16,7}$ & $-1$ & $0$ & $0$ & $R_{7}$   \\
\chunkline
$P_{17,8}$ & $+1$ & $0$ & $0$ & $R_{3}$ &
	& $Q_{17,8}$ & $-1$ & $0$ & $0$ & $R_{7}$ \\
\chunkline
$P_{18,9}$ & $+1$ & $0$ & $0$ & $R_{3}$ & 
	& $Q_{18,9}$ & $-1$ & $0$ & $0$ & $R_{7}$ \\
\chunkline
$U_{4,15}$ & $-1$ & $0$ & $0$ & $R_{7}$ & 
	& $V_{4,15}$ & $+1$ & $0$ & $0$ & $R_{3}$ \\
\chunkline
$U_{5,18}$ & $-1$ & $0$ & $-1$ & $R_{6}+R_{7}$ &
	& $V_{5,18}$ & $+1$ & $0$ & $-1$ & $R_{3}+R_{6}$   \\
\chunkline
$U_{7,11}$ & $-1$ & $0$ & $0$ & $R_{7}$ &
	&  $V_{7,11}$ & $+1$ & $0$ & $0$ & $R_{3}$ \\
\chunkline
$X_{65}$ & $0$ & $0$ & $+2$ & $R_{1}+R_{2}+R_{4}$ &
	& $X_{15,14}$ & $0$ & $0$ & $+2$ & $R_{1}+R_{2}+R_{4}$ \\
\chunkline
$X_{67}$ & $0$ & $0$ & $-1$ & $R_{6}$ &
	& $X_{15,16}$ & $0$ & $0$ & $-1$ & $R_{6}$  \\
\chunkline
$X_{75}$ & $0$ & $-2$ & $+1$ & $R_{4}+R_{5}$ &
	& $X_{16,14}$ & $0$ & $-2$ & $+1$ & $R_{4}+R_{5}$  \\
\chunkline
$X_{89}$ & $0$ & $-2$ & $+2$ & $R_{4}$ &
	& $X_{17,18}$ & $0$ & $-2$ & $+2$ & $R_{4}$ \\
\chunkline
$X_{96}$ & $0$ & $0$ & $-1$ & $R_{5}$ &
	& $X_{18,15}$ & $0$ & $0$ & $-1$ & $R_{5}$ \\
\chunkline
$X_{97}$ & $0$ & $+2$ & $0$ & $R_{1}+R_{2}$ &
	& $X_{18,16}$ & $0$ & $+2$ & $0$ & $R_{1}+R_{2}$ \\
\chunkline \noalign{\vskip\doublerulesep} \chunkline
$U_{1,14}$ & $-1$ & $0$ & $-1$ & $R_{6}+R_{7}$ &
	& $V_{1,14}$ & $+1$ & $0$ & $-1$ & $R_{3}+R_{6}$ \\
\chunkline
$U_{8,12}$ & $-1$ & $+1$ & $0$ & $R_{1}+R_{7}$ &
	&  $V_{8,12}$ & $+1$ & $+1$ & $0$ & $R_{1}+R_{3}$ \\
\chunkline
$X_{18}$ & $0$ & $+1$ & $0$ & $R_{2}$ &
	& $X_{10,17}$ & $0$ & $+1$ & $0$ & $R_{2}$   \\
\chunkline
$X_{21}$ & $0$ & $+1$ & $0$ & $R_{1}$ &
	& $X_{11,10}$ & $0$ & $+1$ & $0$ & $R_{1}$ \\
\chunkline
$X_{28}$ & $0$ & $0$ & $-2$ & $R_{5}+R_{6}$ &
	& $X_{11,17}$ & $0$ & $0$ & $-2$ & $R_{5}+R_{6}$  \\
\chunkline
$X_{32}$ & $0$ & $-1$ & $+2$ & $R_{2}+R_{4}$ &
	& $X_{12,11}$ & $0$ & $-1$ & $+2$ & $R_{2}+R_{4}$  \\
\chunkline
$X_{34}$ & $0$ & $0$ & $-1$ & $R_{6}$ &
	& $X_{12,13}$ & $0$ & $0$ & $-1$ & $R_{6}$  \\
\chunkline
$X_{41}$ & $0$ & $-2$ & $+1$ & $R_{4}+R_{5}$ &
	& $X_{13,10}$ & $0$ & $-2$ & $+1$ & $R_{4}+R_{5}$ \\
\chunkline
$X_{53}$ & $0$ & $0$ & $-1$ & $R_{5}$ &
	& $X_{14,12}$ & $0$ & $0$ & $-1$ & $R_{5}$ \\
\chunkline
$X_{54}$ & $0$ & $+2$ & $0$ & $R_{1}+R_{2}$ &
	& $X_{14,13}$ & $0$ & $+2$ & $0$ & $R_{1}+R_{2}$  \\
\chunkline
\end{tabular}
}
\caption{
Global symmetry charges on the chiral fields of Model A. The $R_a$ refer to the $U(1)_R$ charges summarized in \tref{tab_01}.
 \label{tab_02}}
\end{table}

\subsection{Model B} \label{sec_modelB}

The quiver and toric diagram for Model B, shown in \fref{fig:modelAB}, 
are identical to the ones for Model A.
In the conventions used here, Model B has the same $E$-terms as Model A
but different $J$-terms. 
Its interaction terms are of the following form,
\beal{es02b01}
\resizebox{0.9\textwidth}{!}{$
\begin{array}{rrclcrcl}
& & J & & & & E &  \\
\Lambda_{15}: &
X_{53} \cdot X_{32} \cdot X_{21} &-& X_{54} \cdot X_{41}
& ~~ &
V_{1,14} \cdot Q_{14,5} &-& U_{1,14} \cdot P_{14,5}
\\
\Lambda_{83}: &
X_{34} \cdot X_{41} \cdot X_{18} &-& X_{32} \cdot X_{28}
& ~~ &
V_{8,12} \cdot Q_{12,3} &-& U_{8,12} \cdot P_{12,3}
\\
\Lambda_{46}: &
X_{65} \cdot X_{54} &-& X_{67} \cdot X_{75} \cdot X_{53} \cdot X_{34}
& ~~ &
V_{4,15} \cdot Q_{15,6} &-& U_{4,15} \cdot P_{15,6}
\\
\Lambda_{59}: &
X_{97} \cdot X_{75} &-& X_{96} \cdot X_{65}
& ~~ &
V_{5,18} \cdot Q_{18,9} &-& U_{5,18} \cdot P_{18,9}
\\
\Lambda_{72}: &
X_{28} \cdot X_{89} \cdot X_{96} \cdot X_{67} &-& X_{21} \cdot X_{18} \cdot X_{89} \cdot X_{97}
& ~~ &
V_{7,11} \cdot Q_{11,2} &-& U_{7,11} \cdot P_{11,2}
\\
\Lambda_{10,14}: &
X_{14,12} \cdot X_{12,11} \cdot X_{11,10} &-& X_{14,13} \cdot X_{13,10}
& ~~ &
P_{10,1} \cdot U_{1,14} &-& Q_{10,1} \cdot V_{1,14}
\\
\Lambda_{17,12}: &
X_{12,13} \cdot X_{13,10} \cdot X_{10,17} &-& X_{12,11} \cdot X_{11,17}
& ~~ &
P_{17,8} \cdot U_{8,12} &-& Q_{17,8} \cdot V_{8,12}
\\
\Lambda_{13,15}: &
X_{15,14} \cdot X_{14,13} &-& X_{15,16} \cdot X_{16,14} \cdot X_{14,12} \cdot X_{12,13}
& ~~ &
P_{13,4} \cdot U_{4,15} &-& Q_{13,4} \cdot V_{4,15}
\\
\Lambda_{14,18}: &
X_{18,16} \cdot X_{16,14} &-& X_{18,15} \cdot X_{15,14}
& ~~ &
P_{14,5} \cdot U_{5,18} &-& Q_{14,5} \cdot V_{5,18}
\\
\Lambda_{16,11}: &
X_{11,17} \cdot X_{17,18} \cdot X_{18,15} \cdot X_{15,16} &-& X_{11,10} \cdot X_{10,17} \cdot X_{17,18} \cdot X_{18,16}
& ~~ &
P_{16,7} \cdot U_{7,11} &-& Q_{16,7} \cdot V_{7,11}
\\
\Lambda^{1}_{10,8}: &
V_{8,12} \cdot X_{12,13} \cdot X_{13,10} &-& X_{89} \cdot X_{97} \cdot V_{7,11} \cdot X_{11,10}
& ~~ &
X_{10,17} \cdot Q_{17,8} &-& Q_{10,1} \cdot X_{18}
\\
\Lambda^{1}_{11,1}: &
V_{1,14} \cdot X_{14,12} \cdot X_{12,11} &-& X_{18} \cdot X_{89} \cdot X_{97} \cdot V_{7,11}
& ~~ &
X_{11,10} \cdot Q_{10,1} &-& Q_{11,2} \cdot X_{21}
\\
\Lambda^{1}_{11,8}: &
X_{89} \cdot X_{96} \cdot X_{67} \cdot V_{7,11} &-& V_{8,12} \cdot X_{12,11}
& ~~ &
X_{11,17} \cdot Q_{17,8} &-& Q_{11,2} \cdot X_{28}
\\
\Lambda^{1}_{12,2}: &
X_{21} \cdot V_{1,14} \cdot X_{14,12} &-& X_{28} \cdot V_{8,12}
& ~~ &
X_{12,11} \cdot Q_{11,2} &-& Q_{12,3} \cdot X_{32}
\\
\Lambda^{1}_{12,4}: &
X_{41} \cdot X_{18} \cdot V_{8,12} &-& V_{4,15} \cdot X_{15,16} \cdot X_{16,14} \cdot X_{14,12}
& ~~ &
X_{12,13} \cdot Q_{13,4} &-& Q_{12,3} \cdot X_{34}
\\
\Lambda^{1}_{13,1}: &
X_{18} \cdot V_{8,12} \cdot X_{12,13} &-& V_{1,14} \cdot X_{14,13}
& ~~ &
X_{13,10} \cdot Q_{10,1} &-& Q_{13,4} \cdot X_{41}
\\
\Lambda^{1}_{14,3}: &
X_{32} \cdot X_{21} \cdot V_{1,14} &-& X_{34} \cdot V_{4,15} \cdot X_{15,16} \cdot X_{16,14}
& ~~ &
X_{14,12} \cdot Q_{12,3} &-& Q_{14,5} \cdot X_{53}
\\
\Lambda^{1}_{14,4}: &
V_{4,15} \cdot X_{15,14} &-& X_{41} \cdot V_{1,14}
& ~~ &
X_{14,13} \cdot Q_{13,4} &-& Q_{14,5} \cdot X_{54}
\\
\Lambda^{1}_{15,5}: &
X_{54} \cdot V_{4,15} &-& V_{5,18} \cdot X_{18,15}
& ~~ &
X_{15,14} \cdot Q_{14,5} &-& Q_{15,6} \cdot X_{65}
\\
\Lambda^{1}_{15,7}: &
V_{7,11} \cdot X_{11,17} \cdot X_{17,18} \cdot X_{18,15} &-& X_{75} \cdot X_{53} \cdot X_{34} \cdot V_{4,15}
& ~~ &
X_{15,16} \cdot Q_{16,7} &-& Q_{15,6} \cdot X_{67}
\\
\Lambda^{1}_{16,5}: &
V_{5,18} \cdot X_{18,16} &-& X_{53} \cdot X_{34} \cdot V_{4,15} \cdot X_{15,16}
& ~~ &
X_{16,14} \cdot Q_{14,5} &-& Q_{16,7} \cdot X_{75}
\\
\Lambda^{1}_{17,9}: &
X_{96} \cdot X_{67} \cdot V_{7,11} \cdot X_{11,17} &-& X_{97} \cdot V_{7,11} \cdot X_{11,10} \cdot X_{10,17}
& ~~ &
X_{17,18} \cdot Q_{18,9} &-& Q_{17,8} \cdot X_{89}
\\
\Lambda^{1}_{18,6}: &
X_{67} \cdot V_{7,11} \cdot X_{11,17} \cdot X_{17,18} &-& X_{65} \cdot V_{5,18}
& ~~ &
X_{18,15} \cdot Q_{15,6} &-& Q_{18,9} \cdot X_{96}
\\
\Lambda^{1}_{18,7}: &
X_{75} \cdot V_{5,18} &-& V_{7,11} \cdot X_{11,10} \cdot X_{10,17} \cdot X_{17,18}
& ~~ &
X_{18,16} \cdot Q_{16,7} &-& Q_{18,9} \cdot X_{97}
\\
\Lambda^{2}_{10,8}: &
U_{8,12} \cdot X_{12,13} \cdot X_{13,10} &-& X_{89} \cdot X_{97} \cdot U_{7,11} \cdot X_{11,10}
& ~~ &
P_{10,1} \cdot X_{18} &-& X_{10,17} \cdot P_{17,8}
\\
\Lambda^{2}_{11,1}: &
U_{1,14} \cdot X_{14,12} \cdot X_{12,11} &-& X_{18} \cdot X_{89} \cdot X_{97} \cdot U_{7,11}
& ~~ &
P_{11,2} \cdot X_{21} &-& X_{11,10} \cdot P_{10,1}
\\
\Lambda^{2}_{11,8}: &
X_{89} \cdot X_{96} \cdot X_{67} \cdot U_{7,11} &-& U_{8,12} \cdot X_{12,11}
& ~~ &
P_{11,2} \cdot X_{28} &-& X_{11,17} \cdot P_{17,8}
\\
\Lambda^{2}_{12,2}: &
X_{21} \cdot U_{1,14} \cdot X_{14,12} &-& X_{28} \cdot U_{8,12}
& ~~ &
P_{12,3} \cdot X_{32} &-& X_{12,11} \cdot P_{11,2}
\\
\Lambda^{2}_{12,4}: &
X_{41} \cdot X_{18} \cdot U_{8,12} &-& U_{4,15} \cdot X_{15,16} \cdot X_{16,14} \cdot X_{14,12}
& ~~ &
P_{12,3} \cdot X_{34} &-& X_{12,13} \cdot P_{13,4}
\\
\Lambda^{2}_{13,1}: &
X_{18} \cdot U_{8,12} \cdot X_{12,13} &-& U_{1,14} \cdot X_{14,13}
& ~~ &
P_{13,4} \cdot X_{41} &-& X_{13,10} \cdot P_{10,1}
\\
\Lambda^{2}_{14,3}: &
X_{32} \cdot X_{21} \cdot U_{1,14} &-& X_{34} \cdot U_{4,15} \cdot X_{15,16} \cdot X_{16,14}
& ~~ &
P_{14,5} \cdot X_{53} &-& X_{14,12} \cdot P_{12,3}
\\
\Lambda^{2}_{14,4}: &
U_{4,15} \cdot X_{15,14} &-& X_{41} \cdot U_{1,14}
& ~~ &
P_{14,5} \cdot X_{54} &-& X_{14,13} \cdot P_{13,4}
\\
\Lambda^{2}_{15,5}: &
X_{54} \cdot U_{4,15} &-& U_{5,18} \cdot X_{18,15}
& ~~ &
P_{15,6} \cdot X_{65} &-& X_{15,14} \cdot P_{14,5}
\\
\Lambda^{2}_{15,7}: &
U_{7,11} \cdot X_{11,17} \cdot X_{17,18} \cdot X_{18,15} &-& X_{75} \cdot X_{53} \cdot X_{34} \cdot U_{4,15}
& ~~ &
P_{15,6} \cdot X_{67} &-& X_{15,16} \cdot P_{16,7}
\\
\Lambda^{2}_{16,5}: &
U_{5,18} \cdot X_{18,16} &-& X_{53} \cdot X_{34} \cdot U_{4,15} \cdot X_{15,16}
& ~~ &
P_{16,7} \cdot X_{75} &-& X_{16,14} \cdot P_{14,5}
\\
\Lambda^{2}_{17,9}: &
X_{96} \cdot X_{67} \cdot U_{7,11} \cdot X_{11,17} &-& X_{97} \cdot U_{7,11} \cdot X_{11,10} \cdot X_{10,17}
& ~~ &
P_{17,8} \cdot X_{89} &-& X_{17,18} \cdot P_{18,9}
\\
\Lambda^{2}_{18,6}: &
X_{67} \cdot U_{7,11} \cdot X_{11,17} \cdot X_{17,18} &-& X_{65} \cdot U_{5,18}
& ~~ &
P_{18,9} \cdot X_{96} &-& X_{18,15} \cdot P_{15,6}
\\
\Lambda^{2}_{18,7}: &
X_{75} \cdot U_{5,18} &-& U_{7,11} \cdot X_{11,10} \cdot X_{10,17} \cdot X_{17,18}
& ~~ &
P_{18,9} \cdot X_{97} &-& X_{18,16} \cdot P_{16,7}
\\
\end{array}
$}
~.~
\nn\\
\eea
We obtain the $P$-matrix for Model B using the forward algorithm for brane brick models, giving us, 
\beal{es02b02}
P^B =
\resizebox{0.55\textwidth}{!}{$
\left(
\begin{array}{c|ccccccc|cccc|ccc|ccc|ccc}
& p_1 & p_2 & p_3 & p_6 & p_4 & p_5 & p_7 & q_1 & q_2 & q_3 & q_4 & u_1 & \cdots & u_{42} & w_1 & \cdots & w_{65} & o^{(1)}_1 & \cdots & o^{(19)}_{42} \\
\hline
 P_{10,1} & 0 & 0 & 1 & 0 & 0 & 0 & 0 & 0 & 0 & 0 & 1 & 0 & \cdots & 0 & 0 & \cdots & 1 & 0 & \cdots & 2 \\
 P_{11,2} & 0 & 0 & 1 & 0 & 0 & 0 & 0 & 0 & 0 & 0 & 1 & 0 & \cdots & 1 & 0 & \cdots & 1 & 0 & \cdots & 1 \\
 P_{12,3} & 0 & 0 & 1 & 0 & 0 & 0 & 0 & 0 & 0 & 0 & 1 & 0 & \cdots & 0 & 0 & \cdots & 0 & 0 & \cdots & 0 \\
 P_{13,4} & 0 & 0 & 1 & 0 & 0 & 0 & 0 & 0 & 0 & 0 & 1 & 0 & \cdots & 0 & 0 & \cdots & 1 & 0 & \cdots & 0 \\
 P_{14,5} & 0 & 0 & 1 & 0 & 0 & 0 & 0 & 0 & 0 & 0 & 1 & 0 & \cdots & 0 & 0 & \cdots & 0 & 0 & \cdots & 1 \\
 P_{15,6} & 0 & 0 & 1 & 0 & 0 & 0 & 0 & 0 & 0 & 0 & 1 & 0 & \cdots & 0 & 0 & \cdots & 0 & 0 & \cdots & 0 \\
 P_{16,7} & 0 & 0 & 1 & 0 & 0 & 0 & 0 & 0 & 0 & 0 & 1 & 0 & \cdots & 0 & 0 & \cdots & 0 & 0 & \cdots & 0 \\
 P_{17,8} & 0 & 0 & 1 & 0 & 0 & 0 & 0 & 0 & 0 & 0 & 1 & 0 & \cdots & 0 & 0 & \cdots & 1 & 1 & \cdots & 1 \\
 P_{18,9} & 0 & 0 & 1 & 0 & 0 & 0 & 0 & 0 & 0 & 0 & 1 & 0 & \cdots & 0 & 0 & \cdots & 1 & 0 & \cdots & 1 \\
 Q_{10,1} & 0 & 0 & 0 & 0 & 0 & 0 & 1 & 0 & 0 & 0 & 1 & 0 & \cdots & 0 & 0 & \cdots & 1 & 0 & \cdots & 2 \\
 Q_{11,2} & 0 & 0 & 0 & 0 & 0 & 0 & 1 & 0 & 0 & 0 & 1 & 0 & \cdots & 1 & 0 & \cdots & 1 & 0 & \cdots & 1 \\
 Q_{12,3} & 0 & 0 & 0 & 0 & 0 & 0 & 1 & 0 & 0 & 0 & 1 & 0 & \cdots & 0 & 0 & \cdots & 0 & 0 & \cdots & 0 \\
 Q_{13,4} & 0 & 0 & 0 & 0 & 0 & 0 & 1 & 0 & 0 & 0 & 1 & 0 & \cdots & 0 & 0 & \cdots & 1 & 0 & \cdots & 0 \\
 Q_{14,5} & 0 & 0 & 0 & 0 & 0 & 0 & 1 & 0 & 0 & 0 & 1 & 0 & \cdots & 0 & 0 & \cdots & 0 & 0 & \cdots & 1 \\
 Q_{15,6} & 0 & 0 & 0 & 0 & 0 & 0 & 1 & 0 & 0 & 0 & 1 & 0 & \cdots & 0 & 0 & \cdots & 0 & 0 & \cdots & 0 \\
 Q_{16,7} & 0 & 0 & 0 & 0 & 0 & 0 & 1 & 0 & 0 & 0 & 1 & 0 & \cdots & 0 & 0 & \cdots & 0 & 0 & \cdots & 0 \\
 Q_{17,8} & 0 & 0 & 0 & 0 & 0 & 0 & 1 & 0 & 0 & 0 & 1 & 0 & \cdots & 0 & 0 & \cdots & 1 & 1 & \cdots & 1 \\
 Q_{18,9} & 0 & 0 & 0 & 0 & 0 & 0 & 1 & 0 & 0 & 0 & 1 & 0 & \cdots & 0 & 0 & \cdots & 1 & 0 & \cdots & 1 \\
 U_{1,14} & 1 & 0 & 0 & 0 & 0 & 0 & 1 & 0 & 0 & 1 & 0 & 0 & \cdots & 0 & 0 & \cdots & 0 & 0 & \cdots & 0 \\
 U_{4,15} & 0 & 0 & 0 & 0 & 0 & 0 & 1 & 0 & 0 & 1 & 0 & 0 & \cdots & 0 & 0 & \cdots & 0 & 0 & \cdots & 0 \\
 U_{5,18} & 0 & 0 & 0 & 1 & 0 & 0 & 1 & 1 & 0 & 1 & 0 & 0 & \cdots & 0 & 1 & \cdots & 0 & 0 & \cdots & 0 \\
 U_{7,11} & 0 & 0 & 0 & 0 & 0 & 0 & 1 & 0 & 0 & 1 & 0 & 0 & \cdots & 0 & 0 & \cdots & 0 & 0 & \cdots & 0 \\
 U_{8,12} & 0 & 0 & 0 & 1 & 0 & 0 & 1 & 0 & 1 & 1 & 0 & 0 & \cdots & 0 & 0 & \cdots & 0 & 0 & \cdots & 2 \\
 V_{1,14} & 1 & 0 & 1 & 0 & 0 & 0 & 0 & 0 & 0 & 1 & 0 & 0 & \cdots & 0 & 0 & \cdots & 0 & 0 & \cdots & 0 \\
 V_{4,15} & 0 & 0 & 1 & 0 & 0 & 0 & 0 & 0 & 0 & 1 & 0 & 0 & \cdots & 0 & 0 & \cdots & 0 & 0 & \cdots & 0 \\
 V_{5,18} & 0 & 0 & 1 & 1 & 0 & 0 & 0 & 1 & 0 & 1 & 0 & 0 & \cdots & 0 & 1 & \cdots & 0 & 0 & \cdots & 0 \\
 V_{7,11} & 0 & 0 & 1 & 0 & 0 & 0 & 0 & 0 & 0 & 1 & 0 & 0 & \cdots & 0 & 0 & \cdots & 0 & 0 & \cdots & 0 \\
 V_{8,12} & 0 & 0 & 1 & 1 & 0 & 0 & 0 & 0 & 1 & 1 & 0 & 0 & \cdots & 0 & 0 & \cdots & 0 & 0 & \cdots & 2 \\
 X_{18} & 0 & 0 & 0 & 0 & 0 & 1 & 0 & 1 & 0 & 0 & 0 & 0 & \cdots & 0 & 0 & \cdots & 0 & 1 & \cdots & 0 \\
 X_{21} & 0 & 0 & 0 & 1 & 0 & 0 & 0 & 0 & 1 & 0 & 0 & 1 & \cdots & 0 & 1 & \cdots & 0 & 0 & \cdots & 1 \\
 X_{28} & 1 & 1 & 0 & 0 & 0 & 0 & 0 & 0 & 0 & 0 & 0 & 1 & \cdots & 0 & 1 & \cdots & 0 & 1 & \cdots & 0 \\
 X_{32} & 0 & 0 & 0 & 0 & 1 & 1 & 0 & 1 & 0 & 0 & 0 & 0 & \cdots & 1 & 0 & \cdots & 1 & 1 & \cdots & 2 \\
 X_{34} & 1 & 0 & 0 & 0 & 0 & 0 & 0 & 0 & 0 & 0 & 0 & 0 & \cdots & 0 & 1 & \cdots & 1 & 0 & \cdots & 0 \\
 X_{41} & 0 & 1 & 0 & 0 & 1 & 0 & 0 & 0 & 0 & 0 & 0 & 1 & \cdots & 1 & 0 & \cdots & 0 & 1 & \cdots & 2 \\
 X_{53} & 0 & 1 & 0 & 0 & 0 & 0 & 0 & 0 & 0 & 0 & 0 & 0 & \cdots & 0 & 0 & \cdots & 0 & 1 & \cdots & 0 \\
 X_{54} & 0 & 0 & 0 & 1 & 0 & 1 & 0 & 1 & 1 & 0 & 0 & 0 & \cdots & 0 & 1 & \cdots & 1 & 1 & \cdots & 1 \\
 X_{65} & 1 & 1 & 0 & 0 & 1 & 0 & 0 & 0 & 0 & 0 & 0 & 1 & \cdots & 1 & 0 & \cdots & 0 & 1 & \cdots & 3 \\
 X_{67} & 0 & 0 & 0 & 1 & 0 & 0 & 0 & 1 & 0 & 0 & 0 & 0 & \cdots & 0 & 0 & \cdots & 0 & 0 & \cdots & 3 \\
 X_{75} & 0 & 0 & 0 & 0 & 1 & 1 & 0 & 0 & 1 & 0 & 0 & 1 & \cdots & 1 & 0 & \cdots & 0 & 1 & \cdots & 1 \\
 X_{89} & 0 & 0 & 0 & 0 & 1 & 0 & 0 & 0 & 0 & 0 & 0 & 0 & \cdots & 0 & 0 & \cdots & 0 & 0 & \cdots & 0 \\
 X_{96} & 0 & 0 & 0 & 0 & 0 & 1 & 0 & 0 & 1 & 0 & 0 & 0 & \cdots & 0 & 0 & \cdots & 0 & 1 & \cdots & 0 \\
 X_{97} & 1 & 1 & 0 & 0 & 0 & 0 & 0 & 0 & 0 & 0 & 0 & 0 & \cdots & 0 & 0 & \cdots & 0 & 1 & \cdots & 2 \\
 X_{10,17} & 0 & 0 & 0 & 0 & 0 & 1 & 0 & 1 & 0 & 0 & 0 & 0 & \cdots & 0 & 0 & \cdots & 0 & 0 & \cdots & 1 \\
 X_{11,10} & 0 & 0 & 0 & 1 & 0 & 0 & 0 & 0 & 1 & 0 & 0 & 1 & \cdots & 1 & 1 & \cdots & 0 & 0 & \cdots & 0 \\
 X_{11,17} & 1 & 1 & 0 & 0 & 0 & 0 & 0 & 0 & 0 & 0 & 0 & 1 & \cdots & 1 & 1 & \cdots & 0 & 0 & \cdots & 0 \\
 X_{12,11} & 0 & 0 & 0 & 0 & 1 & 1 & 0 & 1 & 0 & 0 & 0 & 0 & \cdots & 0 & 0 & \cdots & 0 & 1 & \cdots & 1 \\
 X_{12,13} & 1 & 0 & 0 & 0 & 0 & 0 & 0 & 0 & 0 & 0 & 0 & 0 & \cdots & 0 & 1 & \cdots & 0 & 0 & \cdots & 0 \\
 X_{13,10} & 0 & 1 & 0 & 0 & 1 & 0 & 0 & 0 & 0 & 0 & 0 & 1 & \cdots & 1 & 0 & \cdots & 0 & 1 & \cdots & 0 \\
 X_{14,12} & 0 & 1 & 0 & 0 & 0 & 0 & 0 & 0 & 0 & 0 & 0 & 0 & \cdots & 0 & 0 & \cdots & 0 & 1 & \cdots & 1 \\
 X_{14,13} & 0 & 0 & 0 & 1 & 0 & 1 & 0 & 1 & 1 & 0 & 0 & 0 & \cdots & 0 & 1 & \cdots & 0 & 1 & \cdots & 2 \\
 X_{15,14} & 1 & 1 & 0 & 0 & 1 & 0 & 0 & 0 & 0 & 0 & 0 & 1 & \cdots & 1 & 0 & \cdots & 0 & 1 & \cdots & 2 \\
 X_{15,16} & 0 & 0 & 0 & 1 & 0 & 0 & 0 & 1 & 0 & 0 & 0 & 0 & \cdots & 0 & 0 & \cdots & 0 & 0 & \cdots & 3 \\
 X_{16,14} & 0 & 0 & 0 & 0 & 1 & 1 & 0 & 0 & 1 & 0 & 0 & 1 & \cdots & 1 & 0 & \cdots & 0 & 1 & \cdots & 0 \\
 X_{17,18} & 0 & 0 & 0 & 0 & 1 & 0 & 0 & 0 & 0 & 0 & 0 & 0 & \cdots & 0 & 0 & \cdots & 0 & 1 & \cdots & 0 \\
 X_{18,15} & 0 & 0 & 0 & 0 & 0 & 1 & 0 & 0 & 1 & 0 & 0 & 0 & \cdots & 0 & 0 & \cdots & 1 & 1 & \cdots & 1 \\
 X_{18,16} & 1 & 1 & 0 & 0 & 0 & 0 & 0 & 0 & 0 & 0 & 0 & 0 &\cdots & 0 & 0 & \cdots & 1 & 1 & \cdots & 3 \\
\end{array}
\right)
$}
~,~
\eea
where $p_a$ correspond to the extremal GLSM fields.
The columns denoted by $o$ correspond to extra GLSM fields \cite{Franco:2015tna,Franco:2015tya}.
The toric diagram for Model B is obtained from the following coordinate matrix,
\beal{es03a03}
G_t^B=
\resizebox{0.8\textwidth}{!}{$
\left(
\begin{array}{ccccccc|cccc|ccc|ccc|ccc}
p_1 & p_2 & p_3 & p_4 & p_5 & p_6 & p_7 & q_1 & q_2 & q_3 & q_4 & u_1 & \cdots & u_{42} & w_1 & \cdots & w_{65} & o^{(1)}_1 & \cdots & o^{(19)}_{42} \\
\hline
 0 & 1 & -1 & 2 & 2 & 0 & -1 & 1 & 1 & -1 & -1 & 1 & \cdots & 1 & 0 & \cdots & 0 & 3 & \cdots & 2 \\
 -1 & -1 & 0 & 0 & 1 & 1 & 0 & 1 & 1 & 0 & 0 & 0 & \cdots & 0 & 0 & \cdots & 0 & 0 & \cdots & 1 \\
 0 & 0 & 1 & 0 & 0 & 0 & -1 & 0 & 0 & 0 & 0 & 0 & \cdots & 0 & 0 & \cdots & 0 & 0 & \cdots & 0 \\
 \hline
 1 & 1 & 1 & 1 & 1 & 1 & 1 & 1 & 1 & 1 & 1 & 1 & \cdots & 1 & 1 & \cdots & 1 & 2 & \cdots & 4 \\
\end{array}
\right)
$}
~.~
\eea
The toric diagrams for Models A and B agree up to a $GL(3,\mathbb{Z})$ transformation.
Hence, the two models have the same mesonic moduli space $\mathcal{M}^{mes}$.

We also note that Model B has the same enhanced global symmetry as Model A as shown in \eref{es02a04}.
The two models also share the same refined Hilbert series,
\beal{es03a04}
g(x,f_1,f_2,t;\mathcal{M}_{A}^{mes})=g(x,f_1,f_2,t;\mathcal{M}_{B}^{mes}) ~.~
\eea

The $P$-matrices of Models A and B give different compositions of the bifundamental chiral fields in terms of GLSM fields.
Consequently, the global symmetry charges of some chiral fields differ between the two models as shown in \tref{tab_02} and \tref{tab_03}.
The common mesonic moduli space has the same generators and defining relations, but the realizations of these generators
as products of chiral fields differ.
\tref{tab_04} and \tref{tab_04b} compare these realizations
for Models A and B.
\\

\begin{table}[H]
\centering
\newcommand{\chunkline}{\cline{1-5}\cline{7-11}}
\resizebox{0.8\textwidth}{!}{
\begin{tabular}{|c|c|c|c|l| p{1em} |c|c|c|c|l|}
\chunkline
\; & $SU(2)_x$ & $U(1)_{f_1}$ & $U(1)_{f_2}$ & $U(1)_R$ &
   & \; & $SU(2)_x$ & $U(1)_{f_1}$ & $U(1)_{f_2}$ & $U(1)_R$ \\
\chunkline \noalign{\vskip\doublerulesep} \chunkline
$P_{10,1}$ & $+1$ & $0$ & $0$ & $R_{3}$ &
	& $Q_{10,1}$ & $-1$ & $0$ & $0$ & $R_{7}$ \\
\chunkline
$P_{11,2}$ & $+1$ & $0$ & $0$ & $R_{3}$  &
	& $Q_{11,2}$ & $-1$ & $0$ & $0$ & $R_{7}$  \\
\chunkline
$P_{12,3}$ & $+1$ & $0$ & $0$ & $R_{3}$ &
	& $Q_{12,3}$ & $-1$ & $0$ & $0$ & $R_{7}$  \\
\chunkline
$P_{13,4}$ & $+1$ & $0$ & $0$ & $R_{3}$ &
	& $Q_{13,4}$ & $-1$ & $0$ & $0$ & $R_{7}$  \\
\chunkline
$P_{14,5}$ & $+1$ & $0$ & $0$ & $R_{3}$ &
	&  $Q_{14,5}$ & $-1$ & $0$ & $0$ & $R_{7}$ \\
\chunkline
$P_{15,6}$ & $+1$ & $0$ & $0$ & $R_{3}$ &
	& $Q_{15,6}$ & $-1$ & $0$ & $0$ & $R_{7}$  \\
\chunkline
$P_{16,7}$ & $+1$ & $0$ & $0$ & $R_{3}$ &
	& $Q_{16,7}$ & $-1$ & $0$ & $0$ & $R_{7}$   \\
\chunkline
$P_{17,8}$ & $+1$ & $0$ & $0$ & $R_{3}$ &
	& $Q_{17,8}$ & $-1$ & $0$ & $0$ & $R_{7}$ \\
\chunkline
$P_{18,9}$ & $+1$ & $0$ & $0$ & $R_{3}$ & 
	& $Q_{18,9}$ & $-1$ & $0$ & $0$ & $R_{7}$ \\
\chunkline
$U_{4,15}$ & $-1$ & $0$ & $0$ & $R_{7}$ & 
	& $V_{4,15}$ & $+1$ & $0$ & $0$ & $R_{3}$ \\
\chunkline
$U_{5,18}$ & $-1$ & $0$ & $-1$ & $R_{6}+R_{7}$ &
	& $V_{5,18}$ & $+1$ & $0$ & $-1$ & $R_{3}+R_{6}$   \\
\chunkline
$U_{7,11}$ & $-1$ & $0$ & $0$ & $R_{7}$ &
	&  $V_{7,11}$ & $+1$ & $0$ & $0$ & $R_{3}$ \\
\chunkline
$X_{65}$ & $0$ & $0$ & $+2$ & $R_{1}+R_{2}+R_{4}$ &
	& $X_{15,14}$ & $0$ & $0$ & $+2$ & $R_{1}+R_{2}+R_{4}$ \\
\chunkline
$X_{67}$ & $0$ & $0$ & $-1$ & $R_{6}$ &
	& $X_{15,16}$ & $0$ & $0$ & $-1$ & $R_{6}$  \\
\chunkline
$X_{75}$ & $0$ & $-2$ & $+1$ & $R_{4}+R_{5}$ &
	& $X_{16,14}$ & $0$ & $-2$ & $+1$ & $R_{4}+R_{5}$  \\
\chunkline
$X_{89}$ & $0$ & $-2$ & $+2$ & $R_{4}$ &
	& $X_{17,18}$ & $0$ & $-2$ & $+2$ & $R_{4}$ \\
\chunkline
$X_{96}$ & $0$ & $0$ & $-1$ & $R_{5}$ &
	& $X_{18,15}$ & $0$ & $0$ & $-1$ & $R_{5}$ \\
\chunkline
$X_{97}$ & $0$ & $+2$ & $0$ & $R_{1}+R_{2}$ &
	& $X_{18,16}$ & $0$ & $+2$ & $0$ & $R_{1}+R_{2}$ \\
\chunkline \noalign{\vskip\doublerulesep} \chunkline
$U_{1,14}$ & $-1$ & $+1$ & $0$ & $R_{1}+R_{7}$ &
	& $V_{1,14}$ & $+1$ & $+1$ & $0$ & $R_{1}+R_{3}$  \\
\chunkline
$U_{8,12}$ & $-1$ & $0$ & $-1$ & $R_{6}+R_{7}$ &
	& $V_{8,12}$ & $+1$ & $0$ & $-1$ & $R_{3}+R_{6}$ \\
\chunkline
$X_{18}$ & $0$ & $0$ & $-1$ & $R_{5}$ &
	& $X_{10,17}$ & $0$ & $0$ & $-1$ & $R_{5}$  \\
\chunkline
$X_{21}$ & $0$ & $0$ & $-1$ & $R_{6}$ &
	& $X_{11,10}$ & $0$ & $0$ & $-1$ & $R_{6}$  \\
\chunkline
$X_{28}$ & $0$ & $+2$ & $0$ & $R_{1}+R_{2}$ &
	& $X_{11,17}$ & $0$ & $+2$ & $0$ & $R_{1}+R_{2}$  \\
\chunkline
$X_{32}$ & $0$ & $-2$ & $+1$ & $R_{4}+R_{5}$ &
	& $X_{12,11}$ & $0$ & $-2$ & $+1$ & $R_{4}+R_{5}$  \\
\chunkline
$X_{34}$ & $0$ & $+1$ & $0$ & $R_{1}$ &
	& $X_{12,13}$ & $0$ & $+1$ & $0$ & $R_{1}$  \\
\chunkline
$X_{41}$ & $0$ & $-1$ & $+2$ & $R_{2}+R_{4}$ &
	& $X_{13,10}$ & $0$ & $-1$ & $+2$ & $R_{2}+R_{4}$  \\
\chunkline
$X_{53}$ & $0$ & $+1$ & $0$ & $R_{2}$ &
	& $X_{14,12}$ & $0$ & $+1$ & $0$ & $R_{2}$   \\
\chunkline
$X_{54}$ & $0$ & $0$ & $-2$ & $R_{5}+R_{6}$ &
	& $X_{14,13}$ & $0$ & $0$ & $-2$ & $R_{5}+R_{6}$  \\
\chunkline
\end{tabular}
}
\caption{
Global symmetry charges on the chiral fields of Model B. The $R_a$ refer to the $U(1)_R$ charges summarized in \tref{tab_01}.
 \label{tab_03}}
\end{table}

\begin{figure}[H]
\centering
\includegraphics[height=0.33\textheight]{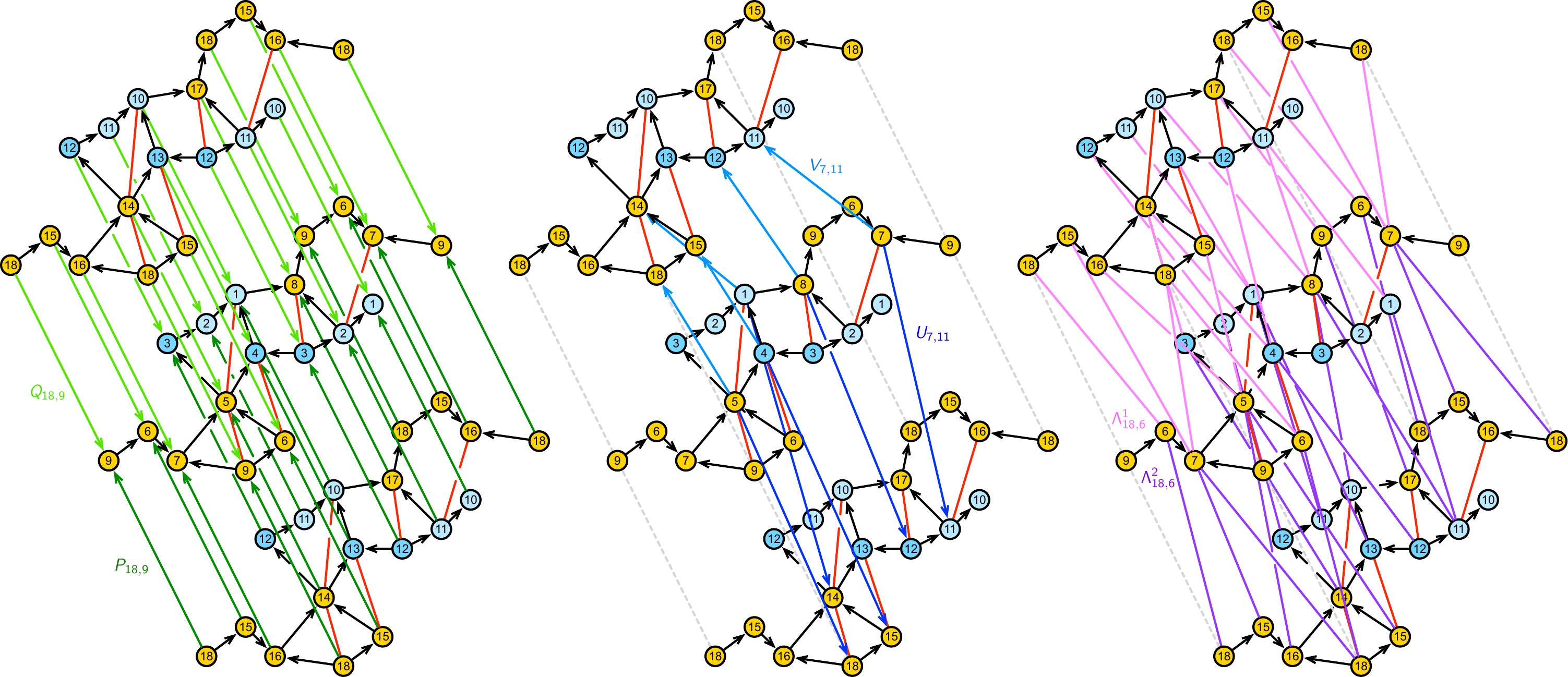}
\caption{The periodic quiver of Model B.}
\label{fig:periodic_quiver_B}
\end{figure}

\begingroup
\small
\setlength{\tabcolsep}{5pt}
\renewcommand{\arraystretch}{1.08}
\captionsetup{position=bottom}
\begin{table}[H]
\centering
\resizebox{0.74\textwidth}{!}{%
\begin{tabular}{@{}l p{0.84\textwidth}@{}}
\; & Generator \\ \hline
GLSM & $p_1^{2} p_2^{2} p_3^{2} p_4 q_{(2)} u w$ \\*
Model A & $V_{7,11} X_{11,10} X_{10,17} X_{17,18} P_{18,9} X_{97}$ \\*
Model B & $V_{7,11} X_{11,17} X_{17,18} P_{18,9} X_{97}$ \\ \hline
GLSM & $p_1^{2} p_2^{2} p_3 p_4 p_7 q_{(2)} u w$ \\*
Model A & $V_{7,11} X_{11,10} X_{10,17} X_{17,18} Q_{18,9} X_{97}$ \\*
Model B & $V_{7,11} X_{11,17} X_{17,18} Q_{18,9} X_{97}$ \\ \hline
GLSM & $p_1^{2} p_2^{2} p_4 p_7^{2} q_{(2)} u w$ \\*
Model A & $U_{7,11} X_{11,10} X_{10,17} X_{17,18} Q_{18,9} X_{97}$ \\*
Model B & $U_{7,11} X_{11,17} X_{17,18} Q_{18,9} X_{97}$ \\ \hline
GLSM & $p_1 p_2 p_3^{2} p_4 p_5 p_6 q_{(1)} q_{(2)} u w$ \\*
Model A & $V_{7,11} X_{11,17} X_{17,18} P_{18,9} X_{97}$ \\*
Model B & $V_{7,11} X_{11,10} X_{10,17} X_{17,18} P_{18,9} X_{97}$ \\ \hline
GLSM & $p_1 p_2 p_3 p_4 p_5 p_6 p_7 q_{(1)} q_{(2)} u w$ \\*
Model A & $V_{7,11} X_{11,17} X_{17,18} Q_{18,9} X_{97}$ \\*
Model B & $V_{7,11} X_{11,10} X_{10,17} X_{17,18} Q_{18,9} X_{97}$ \\ \hline
GLSM & $p_1 p_2 p_4 p_5 p_6 p_7^{2} q_{(1)} q_{(2)} u w$ \\*
Model A & $U_{7,11} X_{11,17} X_{17,18} Q_{18,9} X_{97}$ \\*
Model B & $U_{7,11} X_{11,10} X_{10,17} X_{17,18} Q_{18,9} X_{97}$ \\ \hline
GLSM & $p_3^{2} p_4 p_5^{2} p_6^{2} q_{(1)}^{2} q_{(2)} u w$ \\*
Model A & $X_{67} V_{7,11} X_{11,17} X_{17,18} P_{18,9} X_{96}$ \\*
Model B & $X_{67} V_{7,11} X_{11,10} X_{10,17} X_{17,18} P_{18,9} X_{96}$ \\ \hline
GLSM & $p_3 p_4 p_5^{2} p_6^{2} p_7 q_{(1)}^{2} q_{(2)} u w$ \\*
Model A & $X_{67} V_{7,11} X_{11,17} X_{17,18} Q_{18,9} X_{96}$ \\*
Model B & $X_{67} V_{7,11} X_{11,10} X_{10,17} X_{17,18} Q_{18,9} X_{96}$ \\ \hline
GLSM & $p_4 p_5^{2} p_6^{2} p_7^{2} q_{(1)}^{2} q_{(2)} u w$ \\*
Model A & $X_{67} U_{7,11} X_{11,17} X_{17,18} Q_{18,9} X_{96}$ \\*
Model B & $X_{67} U_{7,11} X_{11,10} X_{10,17} X_{17,18} Q_{18,9} X_{96}$ \\ \hline
GLSM & $p_1^{2} p_2^{3} p_4^{3} p_5^{2} q_{(1)} u^{2} w$ \\*
Model A & $X_{10,17} X_{17,18} X_{18,16} X_{16,14} X_{14,13} X_{13,10}$ \\*
Model B & $X_{11,17} X_{17,18} X_{18,16} X_{16,14} X_{14,12} X_{12,11}$ \\ \hline
GLSM & $p_1 p_2^{2} p_4^{3} p_5^{3} p_6 q_{(1)}^{2} u^{2} w$ \\*
Model A & $X_{11,17} X_{17,18} X_{18,16} X_{16,14} X_{14,12} X_{12,11}$ \\*
Model B & $X_{11,17} X_{17,18} X_{18,15} X_{15,16} X_{16,14} X_{14,12} X_{12,11}$ \\ \hline
GLSM & $p_2 p_4^{3} p_5^{4} p_6^{2} q_{(1)}^{3} u^{2} w$ \\*
Model A & $X_{11,17} X_{17,18} X_{18,15} X_{15,16} X_{16,14} X_{14,12} X_{12,11}$ \\*
Model B & $X_{10,17} X_{17,18} X_{18,15} X_{15,16} X_{16,14} X_{14,13} X_{13,10}$ \\ \hline
GLSM & $p_1^{2} p_2 p_3^{6} p_6^{2} q_{(1)} q_{(2)}^{3} u w^{2} $ \\*
Model A & $V_{4,15} X_{15,16} P_{16,7} V_{7,11} X_{11,10} X_{10,17} P_{17,8} V_{8,12} X_{12,13} P_{13,4}$ \\*
Model B & $V_{4,15} X_{15,16} P_{16,7} V_{7,11} X_{11,17} P_{17,8} V_{8,12} X_{12,13} P_{13,4}$ \\ \hline
\end{tabular}%
}
\caption{
Generators of the common mesonic moduli space, together with the corresponding chiral field product for each model. 
We set all extra GLSM fields to $1$ and use $q_{(1)}=q_1q_2$, $q_{(2)}=q_3q_4$, $u=\prod_{b=1}^{42}u_b$ and $w=\prod_{b=1}^{65}w_b$.
\textbf{(Part 1/2)}
}
\label{tab_04}
\end{table}
\endgroup

\begingroup
\small
\setlength{\tabcolsep}{5pt}
\renewcommand{\arraystretch}{1.08}
\captionsetup{position=bottom}
\begin{table}[H]
\centering
\resizebox{0.74\textwidth}{!}{%
\begin{tabular}{@{}l p{0.84\textwidth}@{}}
\; & Generator \\ \hline
GLSM & $p_1^{2} p_2 p_3^{5} p_6^{2} p_7 q_{(1)} q_{(2)}^{3} u w^{2}$ \\*
Model A & $V_{4,15} X_{15,16} P_{16,7} V_{7,11} X_{11,10} X_{10,17} P_{17,8} V_{8,12} X_{12,13} Q_{13,4}$ \\*
Model B & $V_{4,15} X_{15,16} P_{16,7} V_{7,11} X_{11,17} P_{17,8} V_{8,12} X_{12,13} Q_{13,4}$ \\ \hline
GLSM & $p_1^{2} p_2 p_3^{4} p_6^{2} p_7^{2} q_{(1)} q_{(2)}^{3} u w^{2}$ \\*
Model A & $V_{1,14} Q_{14,5} V_{5,18} P_{18,9} X_{97} V_{7,11} Q_{11,2} X_{21}$ \\*
Model B & $V_{1,14} Q_{14,5} V_{5,18} P_{18,9} X_{97} V_{7,11} Q_{11,2} X_{21}$ \\ \hline
GLSM & $p_1^{2} p_2 p_3^{3} p_6^{2} p_7^{3} q_{(1)} q_{(2)}^{3} u w^{2}$ \\*
Model A & $V_{4,15} X_{15,16} Q_{16,7} V_{7,11} X_{11,10} X_{10,17} Q_{17,8} V_{8,12} X_{12,13} Q_{13,4}$ \\*
Model B & $V_{4,15} X_{15,16} Q_{16,7} V_{7,11} X_{11,17} Q_{17,8} V_{8,12} X_{12,13} Q_{13,4}$ \\ \hline
GLSM & $p_1^{2} p_2 p_3^{2} p_6^{2} p_7^{4} q_{(1)} q_{(2)}^{3} u w^{2}$ \\*
Model A & $V_{4,15} X_{15,16} Q_{16,7} V_{7,11} X_{11,10} X_{10,17} Q_{17,8} U_{8,12} X_{12,13} Q_{13,4}$ \\*
Model B & $V_{4,15} X_{15,16} Q_{16,7} V_{7,11} X_{11,17} Q_{17,8} U_{8,12} X_{12,13} Q_{13,4}$ \\ \hline
GLSM & $p_1^{2} p_2 p_3 p_6^{2} p_7^{5} q_{(1)} q_{(2)}^{3} u w^{2}$ \\*
Model A & $V_{4,15} X_{15,16} Q_{16,7} U_{7,11} X_{11,10} X_{10,17} Q_{17,8} U_{8,12} X_{12,13} Q_{13,4}$ \\*
Model B & $V_{4,15} X_{15,16} Q_{16,7} U_{7,11} X_{11,17} Q_{17,8} U_{8,12} X_{12,13} Q_{13,4}$ \\ \hline
GLSM & $p_1^{2} p_2 p_6^{2} p_7^{6} q_{(1)} q_{(2)}^{3} u w^{2}$ \\*
Model A & $U_{4,15} X_{15,16} Q_{16,7} U_{7,11} X_{11,10} X_{10,17} Q_{17,8} U_{8,12} X_{12,13} Q_{13,4}$ \\*
Model B & $U_{4,15} X_{15,16} Q_{16,7} U_{7,11} X_{11,17} Q_{17,8} U_{8,12} X_{12,13} Q_{13,4}$ \\ \hline
GLSM & $p_1 p_3^{6} p_5 p_6^{3} q_{(1)}^{2} q_{(2)}^{3} u w^{2}$ \\*
Model A & $V_{4,15} X_{15,16} P_{16,7} V_{7,11} X_{11,17} P_{17,8} V_{8,12} X_{12,13} P_{13,4}$ \\*
Model B & $V_{4,15} X_{15,16} P_{16,7} V_{7,11} X_{11,10} X_{10,17} P_{17,8} V_{8,12} X_{12,13} P_{13,4}$ \\ \hline
GLSM & $p_1 p_3^{5} p_5 p_6^{3} p_7 q_{(1)}^{2} q_{(2)}^{3} u w^{2}$ \\*
Model A & $V_{4,15} X_{15,16} P_{16,7} V_{7,11} X_{11,17} P_{17,8} V_{8,12} X_{12,13} Q_{13,4}$ \\*
Model B & $V_{4,15} X_{15,16} P_{16,7} V_{7,11} X_{11,10} X_{10,17} P_{17,8} V_{8,12} X_{12,13} Q_{13,4}$ \\ \hline
GLSM & $p_1 p_3^{4} p_5 p_6^{3} p_7^{2} q_{(1)}^{2} q_{(2)}^{3} u w^{2}$ \\*
Model A & $V_{1,14} Q_{14,5} V_{5,18} P_{18,9} X_{96} X_{67} V_{7,11} Q_{11,2} X_{21}$ \\*
Model B & $V_{1,14} Q_{14,5} V_{5,18} P_{18,9} X_{96} X_{67} V_{7,11} Q_{11,2} X_{21}$ \\ \hline
GLSM & $p_1 p_3^{3} p_5 p_6^{3} p_7^{3} q_{(1)}^{2} q_{(2)}^{3} u w^{2}$ \\*
Model A & $V_{4,15} X_{15,16} Q_{16,7} V_{7,11} X_{11,17} Q_{17,8} V_{8,12} X_{12,13} Q_{13,4}$ \\*
Model B & $V_{4,15} X_{15,16} Q_{16,7} V_{7,11} X_{11,10} X_{10,17} Q_{17,8} V_{8,12} X_{12,13} Q_{13,4}$ \\ \hline
GLSM & $p_1 p_3^{2} p_5 p_6^{3} p_7^{4} q_{(1)}^{2} q_{(2)}^{3} u w^{2}$ \\*
Model A & $V_{4,15} X_{15,16} Q_{16,7} V_{7,11} X_{11,17} Q_{17,8} U_{8,12} X_{12,13} Q_{13,4}$ \\*
Model B & $V_{4,15} X_{15,16} Q_{16,7} V_{7,11} X_{11,10} X_{10,17} Q_{17,8} U_{8,12} X_{12,13} Q_{13,4}$ \\ \hline
GLSM & $p_1 p_3 p_5 p_6^{3} p_7^{5} q_{(1)}^{2} q_{(2)}^{3} u w^{2}$ \\*
Model A & $V_{4,15} X_{15,16} Q_{16,7} U_{7,11} X_{11,17} Q_{17,8} U_{8,12} X_{12,13} Q_{13,4}$ \\*
Model B & $V_{4,15} X_{15,16} Q_{16,7} U_{7,11} X_{11,10} X_{10,17} Q_{17,8} U_{8,12} X_{12,13} Q_{13,4}$ \\ \hline
GLSM & $p_1 p_5 p_6^{3} p_7^{6} q_{(1)}^{2} q_{(2)}^{3} u w^{2}$ \\*
Model A & $U_{4,15} X_{15,16} Q_{16,7} U_{7,11} X_{11,17} Q_{17,8} U_{8,12} X_{12,13} Q_{13,4}$ \\*
Model B & $U_{4,15} X_{15,16} Q_{16,7} U_{7,11} X_{11,10} X_{10,17} Q_{17,8} U_{8,12} X_{12,13} Q_{13,4}$ \\ \hline
\end{tabular}%
}
\caption{
Generators of the common mesonic moduli space, together with the corresponding chiral field product for each model. 
We set all extra GLSM fields to $1$ and use $q_{(1)}=q_1q_2$, $q_{(2)}=q_3q_4$, $u=\prod_{b=1}^{42}u_b$ and $w=\prod_{b=1}^{65}w_b$.
\textbf{(Part 2/2)}
}
\label{tab_04b}
\end{table}
\endgroup

\subsection{Quiver-Invariant Triality Sequence}
\label{sec:triality-chain}

\begin{figure}[H]
\centering
\includegraphics[width=1\textwidth]{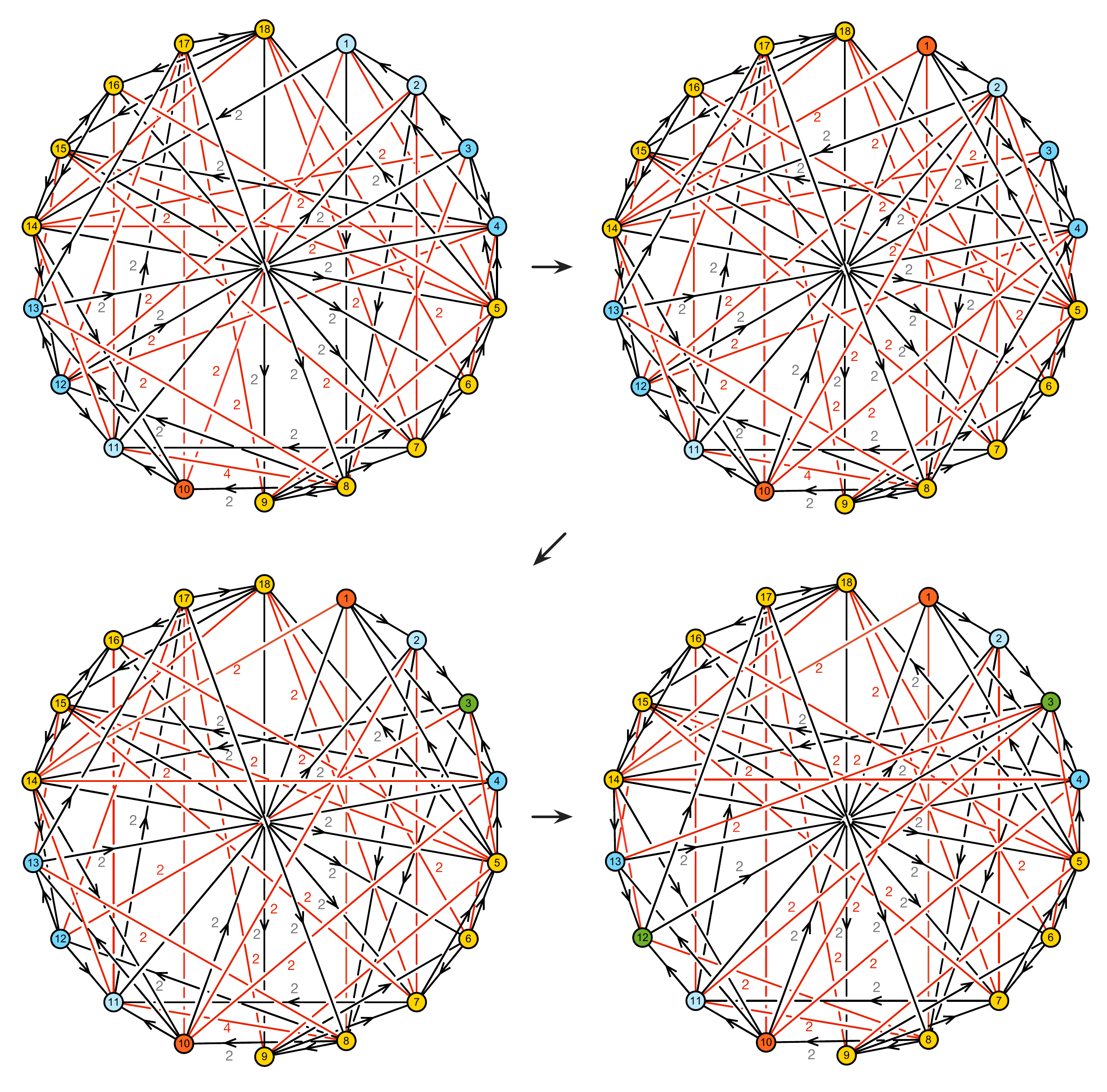}
\caption{Successive quivers obtained from Model A by $\tau_{10}$, $\tau_1$, $\tau_3^{-1}$ and $\tau_{12}^{-1}$, in reading order. After relabelling of gauge nodes $1\leftrightarrow3$ and $10\leftrightarrow12$, the final quiver for Model B agrees with the one for Model A. Red gauge nodes correspond to triality mutation and green gauge nodes correspond to inverse triality mutation. The remaining gauge nodes are colored as in the periodic quivers for Models A and B. }
\label{fig_sequence}
\end{figure}

The periodic quivers distinguish Models A and B 
even though their regular quivers agree. 
The sequence, as illustrated in \fref{fig_sequence}, acts first with triality at nodes $10$ and $1$, 
and then with inverse triality at nodes $3$ and $12$ as follows, 
\beal{es04a01}
\mathcal{T}_A\xrightarrow{\tau_{10}}\mathcal{T}_1
\xrightarrow{\tau_1}\mathcal{T}_2
\xrightarrow{\tau_3^{-1}}\mathcal{T}_3
\xrightarrow{\pi\circ\tau_{12}^{-1}}\mathcal{T}_B
~,~
\eea
where $\mathcal{T}_A$ and $\mathcal{T}_B$ denote Models A and B,
respectively, and $\mathcal{T}_{1,2,3}$ denote the intermediate brane brick models.
The permutation $\pi$ exchanges $1\leftrightarrow3$ and $10\leftrightarrow12$, leaving all other gauge node labels fixed.
At each step in \eref{es04a01}, the mesons and transformed interaction plaquettes must be included, and massive chiral-Fermi pairs must be integrated out.
The sequence restores the underlying quiver up to the relabelling $\pi$ and maps the interaction terms of Model A to those of Model B.
In the displayed conventions, the endpoint $E$-terms coincide, while the $J$-terms differ.
The different $P$-matrices and realizations of the mesonic generators described in sections \sref{sec_modelA} and \sref{sec_modelB} make the distinction between the two models explicit.
\\

\subsection{Elliptic Genus and Quiver-Invariant Triality}
\label{sec:elliptic-genus}

Triality predicts the equality of protected infrared observables, including the flavored elliptic genus, for $2d$ $(0,2)$ theories related by triality.
A direct comparison of the elliptic genus using the supersymmetric localization formula of \cite{Benini:2013nda,Benini:2013xpa}, however, is computationally challenging for the two brane brick models considered in this work due to their large number of gauge groups. 
For abelian brane brick models, an alternative is provided by the geometric prescription reviewed in section \sref{sec021}.
The resulting modified elliptic genus depends on the mesonic moduli space and the mesonic flavor charges of the extremal GLSM fields, rather than on the charge assignments of the elementary chiral fields appearing in the $2d$ $(0,2)$ theory. 

We first relate the geometric fugacities $\mathbf{t}=(t_1,t_2,t_3,t_4)$ to the mesonic flavor fugacities $\mathbf{x}=(x,f_1,f_2)$, associated with the mesonic flavor symmetry $SU(2)_x\times U(1)_{f_1}\times U(1)_{f_2}$ for Models A and B.
Following the approach developed for toric Calabi-Yau 3-folds and $4d$ $\mathcal{N}=1$ gauge theories \cite{Butti:2006au,Forcella:2007wk,Butti:2007jv}, 
let $v_a=(v_a^1,v_a^2,v_a^3,v_a^4)\in\mathbb{Z}^4$ denote the columns of the $G_t$ matrix corresponding to the $7$ extremal GLSM fields $p_a$ as illustrated in \fref{fig:modelAB} and given in \eref{es02a03} and \eref{es03a03}.
We denote their mesonic flavor charge vectors by $\rho_{\mathbf{x}}(p_a)$ as given in \tref{tab_01}. 
The geometric fugacities are then given by,
\beal{es07a01}
t_i=\prod_{a=1}^{7}
\left(\mathbf{x}^{\rho_{\mathbf{x}}(p_a)}\right)^{v_a^i}\,,
\qquad i=1,\ldots,4
~.~
\eea
Using the charge assignments in \tref{tab_01} and the toric data in \eqref{es02a03} and \eqref{es03a03}, 
we obtain,
\beal{es07a02}
&
t_1 =f_1^{-3}f_2^{2}~,~ 
t_2=f_1^{-2}f_2^{-2}~,~
t_3 =x^{2}~,~ 
t_4=1~.~
&
\eea
The value for $t_4=1$, also observed in \cite{Franco:2017cjj}, follows from the Calabi-Yau hyperplane condition $v_a^4=1$
and the vanishing total charge $\sum_{a=1}^{7}\rho_{\mathbf{x}}(p_a)=(0,0,0)$ under the mesonic flavor symmetries.
The auxiliary grading fugacity $t$ used in the Hilbert series
is not included in this flavor fugacity map.

We next choose a triangulation $\mathcal{T}$ of the toric diagram $\Delta$ in \fref{fig:modelAB} into eighteen tetrahedra.
In addition to $v_1,\ldots,v_7$, we also introduce,
\beal{es07a05}
&
 v_8=(1,1,0,1) ~,~ v_9=(-1,0,0,1) ~,~
 v_{10}=(1,0,0,1) ~,~ v_{11}=(0,0,0,1)
 ~,~
 &
\eea
corresponding, respectively, to the points associated with $q_{1,2}$, $q_{3,4}$, $u_a$ and $w_b$ in \fref{fig:modelAB}.
We take then,
\beal{es07a06}
\mathcal{T}=\bigl\{(i,j,k,\ell)\mid
\ell\in\{3,7\},\ (i,j,k)\in\mathcal{B}\bigr\}
~,~
\eea
where $\mathcal{B}$ is the following triangulation of the $v^3=0$, $v^4=1$ slice,
\beal{es07a07}
\mathcal{B}=
&&
\bigl\{(9,6,11),\ (9,11,1),\ (6,11,8),\ (8,11,10),\ (11,1,2),
\nn\\
&&
\hspace{1cm}
 (11,2,10), \ (5,8,10),\ (5,10,2),\ (4,5,2)\bigr\}
~.~
\eea

When evaluating the geometric formula, we order the vertices of each $\sigma\in\mathcal{T}$ so that
$\det(v_{\sigma_1},v_{\sigma_2},v_{\sigma_3},v_{\sigma_4})=+1$.
The geometric prescription reviewed in section \sref{sec021} then gives the modified elliptic genus as a sum over the tetrahedra as follows,
\beal{es07a08}
\mathcal{I}(q;x,f_1,f_2)
=-i\eta(q)^3\sum_{\sigma\in\mathcal{T}}
\frac{\prod_{n=1}^{3}\theta_1(q,z_n^{\sigma})}
{\prod_{d=1}^{4}\theta_1(q,y_d^{\sigma})}
~,~
\eea
where the arguments $y_d^{\sigma}$ and $z_n^{\sigma}$ are determined by the dual cone associated with each tetrahedron, 
with the flavor fugacities related to the geometric ones by \eref{es07a02} above.

For the chosen triangulation, the result can be written in a form manifestly invariant under the $SU(2)_x$ Weyl reflection
$x\leftrightarrow x^{-1}$ as follows,
\beal{es07a10}
 \mathcal{I}(q;x,f_1,f_2)
 =-i\eta(q)^3\Bigl[
 \mathcal{F}(x;f_1,f_2)+\mathcal{F}(x^{-1};f_1,f_2)
 \Bigr]
 ~.~
\eea
Using the abbreviation $[z]_q\equiv\theta_1(q,z)$,
the function $\mathcal{F}$ is given by,
\beal{es07a11}
&&
\mathcal{F}(x;f_1,f_2)
=
\frac{
[x^{2}f_{1}^{5}]_{q}
[x^{-2}f_{1}^{-2}f_{2}^{-2}]_{q}
[x^{-4}f_{1}^{-3}f_{2}^{2}]_{q}
}{
[x^{-2}]_{q}
[x^{4}f_{1}^{5}]_{q}
[f_{1}^{-2}f_{2}^{-2}]_{q}
[x^{-2}f_{1}^{-3}f_{2}^{2}]_{q}
}
+
\frac{
[x^{-4}f_{1}^{-5}]_{q}
[x^{2}f_{1}^{3}f_{2}^{-2}]_{q}
[x^{-2}f_{1}^{2}f_{2}^{2}]_{q}
}{
[x^{-2}]_{q}
[x^{-2}f_{1}^{-5}]_{q}
[x^{4}f_{1}^{3}f_{2}^{-2}]_{q}
[f_{1}^{2}f_{2}^{2}]_{q}
}
\nn\\
&&
\hspace{0.5cm}
+
\frac{
[x^{2}f_{1}f_{2}^{-4}]_{q}
[x^{-4}f_{1}^{-3}f_{2}^{2}]_{q}
[x^{-2}f_{1}^{2}f_{2}^{2}]_{q}
}{
[x^{-2}]_{q}
[x^{4}f_{1}f_{2}^{-4}]_{q}
[x^{-2}f_{1}^{-3}f_{2}^{2}]_{q}
[f_{1}^{2}f_{2}^{2}]_{q}
}
+
\frac{
[f_{1}f_{2}^{-4}]_{q}
[x^{-4}f_{1}^{-3}f_{2}^{2}]_{q}
[f_{1}^{2}f_{2}^{2}]_{q}
}{
[x^{-2}]_{q}
[x^{2}f_{1}f_{2}^{-4}]_{q}
[x^{-2}f_{1}^{-3}f_{2}^{2}]_{q}
[x^{2}f_{1}^{2}f_{2}^{2}]_{q}
}
\nn\\
&&
\hspace{0.5cm}
+
\frac{
[x^{-2}f_{1}^{-2}f_{2}^{-2}]_{q}
[x^{2}f_{1}^{3}f_{2}^{-2}]_{q}
[x^{-4}f_{1}^{-1}f_{2}^{4}]_{q}
}{
[x^{-2}]_{q}
[f_{1}^{-2}f_{2}^{-2}]_{q}
[x^{4}f_{1}^{3}f_{2}^{-2}]_{q}
[x^{-2}f_{1}^{-1}f_{2}^{4}]_{q}
}
+
\frac{
[f_{1}^{5}]_{q}
[f_{1}^{-2}f_{2}^{-2}]_{q}
[x^{-4}f_{1}^{-3}f_{2}^{2}]_{q}
}{
[x^{-2}]_{q}
[x^{2}f_{1}^{5}]_{q}
[x^{2}f_{1}^{-2}f_{2}^{-2}]_{q}
[x^{-2}f_{1}^{-3}f_{2}^{2}]_{q}
}
\nn\\
&&
\hspace{0.5cm}
+
\frac{
[x^{2}f_{1}f_{2}^{-4}]_{q}
[x^{-6}f_{1}^{-3}f_{2}^{2}]_{q}
[f_{1}^{2}f_{2}^{2}]_{q}
}{
[x^{-2}]_{q}
[x^{4}f_{1}f_{2}^{-4}]_{q}
[x^{-4}f_{1}^{-3}f_{2}^{2}]_{q}
[x^{2}f_{1}^{2}f_{2}^{2}]_{q}
}
+
\frac{
[x^{8}f_{1}^{4}f_{2}^{-6}]_{q}
[x^{-6}f_{1}^{-3}f_{2}^{2}]_{q}
[x^{-6}f_{1}^{-1}f_{2}^{4}]_{q}
}{
[x^{-2}]_{q}
[x^{10}f_{1}^{4}f_{2}^{-6}]_{q}
[x^{-4}f_{1}^{-3}f_{2}^{2}]_{q}
[x^{-4}f_{1}^{-1}f_{2}^{4}]_{q}
}
\nn\\
&&
\hspace{0.5cm}
+
\frac{
[x^{4}f_{1}f_{2}^{-4}]_{q}
[x^{4}f_{1}^{3}f_{2}^{-2}]_{q}
[x^{-12}f_{1}^{-4}f_{2}^{6}]_{q}
}{
[x^{-2}]_{q}
[x^{6}f_{1}f_{2}^{-4}]_{q}
[x^{6}f_{1}^{3}f_{2}^{-2}]_{q}
[x^{-10}f_{1}^{-4}f_{2}^{6}]_{q}
}
~.~
\eea
The $9$ terms in $\mathcal{F}(x^{-1};f_1,f_2)$ are obtained from those above by replacing $x$ with $x^{-1}$, reproducing all eighteen contributions of the triangulation in \eref{es07a06}.
Since Models A and B have identical toric data and mesonic flavor charges on extremal GLSM fields, 
the geometric prescription gives the same modified elliptic genus for both Models A and B.
\\

\subsection{From Quiver-Invariant Duality to Triality via Orbifold Reduction}
\label{sec:parent-reduction}

Models A and B are obtained by orbifold reduction of the brane tiling pair studied in \cite{Kho:2026geo}.
The parent superpotentials $W_A$ and $W_B$ are given in \cite{Kho:2026geo}.
For both parent $4d$ $\mathcal{N}=1$ theories, we identify a common perfect matching given by, 
\beal{es06a10}
p_0=\{\mathcal{X}_{15},\mathcal{X}_{83},\mathcal{X}_{46},
\mathcal{X}_{59},\mathcal{X}_{72}\}
~,~
\eea
which contains exactly one chiral field from each superpotential term.
We perform orbifold reduction with $k=2$ layers and the sign vector $s=(-1,+1)$ following the layer convention in \eref{es01a61}.

We label the gauge nodes in the first layer by $i=1,\ldots,9$ and those in the second layer by $i+9$.
Each of the $14$ parent chiral fields outside $p_0$ gives rise to one horizontal chiral field in each layer.
In contrast, each parent chiral field $\mathcal{X}_{ij}\in p_0$ gives rise to two interlayer chiral fields denoted by $U_{i,j+9}$ and $V_{i,j+9}$ both directed from node $i$ to node $j+9$.
The parent vector multiplets give rise to the bifundamental chiral fields $P_{i+9,i}$ and $Q_{i+9,i}$.
Furthermore, the chiral fields in $p_0$ give rise to horizontal Fermi fields, while the remaining parent chiral fields give rise to interlayer Fermi fields.
The resulting brane brick models therefore share the same matter content,
\beal{es06a11}
(G,n^\chi,n^F)=(18,56,38)\,,
\qquad n^\chi-n^F=18
~.~
\eea
We note here that the local non-abelian anomaly cancellation condition is satisfied at every gauge node.
The lifted plaquettes give rise to the $J$- and $E$-terms in \eref{es02a01} and \eref{es02b01},
which satisfy the vanishing trace condition.
The tilting mechanism relating the parent brane tilings in \cite{Kho:2026geo} is also visible after orbifold reduction in the periodic quivers of Models A and B, shown in \fref{fig:periodic_quiver_A} and \fref{fig:periodic_quiver_B}.
The blue gauge nodes $1,2,3,4$ and $10,11,12,13$ are the copies, in the two layers, of the parent gauge nodes $1,2,3,4$ associated with the faces involved in the tilting mutation on the corresponding brane tilings.
The rearrangement relating the two periodic quivers thus exhibits the lift of the parent tilting mechanism to
brane brick models related by quiver-invariant triality.

For the quiver-invariant pair of parent brane tilings, orbifold reduction with common data $(p_0,k,s)$ yields brane brick models
with identical quivers and $E$-terms, while the distinct parent superpotentials give rise only to differences in the $J$-terms.
In general, identical parent quivers do not guarantee
the existence of a common perfect matching or identical mesonic moduli spaces after orbifold reduction.
For the present pair, these properties are verified explicitly
through the $J$- and $E$-terms and the corresponding toric diagrams obtained via the forward algorithm.
Moreover, the triality sequence presented in section \sref{sec:triality-chain}
further confirms that the resulting brane brick models share the same mesonic moduli space and correspond to the same toric Calabi-Yau 4-fold.
\\

\section{Discussion and Conclusions \label{sec05}}

In this work, we have studied an explicit example of quiver-invariant triality between brane brick models describing $2d$ $(0,2)$ gauge theories.
Models A and B share the same quiver, with $18$ gauge nodes, $56$ chiral multiplets and $38$ Fermi multiplets, but differ in their $J$-terms. Their $E$-terms coincide in our conventions.
Consequently, their periodic quivers on $T^3$ are distinct, despite having the same underlying quiver. 
We have exhibited a sequence of $2$ triality and $2$ inverse triality transformations that connects the two models.
While the quiver changes at intermediate steps, the final quiver agrees with the initial one up to a relabelling of gauge nodes, whereas the $J$- and $E$-terms are transformed into those of Model B. 
This provides a concrete realization of quiver-invariant triality in brane brick models.

Using the forward algorithm, we have shown that Models A and B have the same mesonic moduli space described by a common toric Calabi-Yau 4-fold.
They also share the enhanced mesonic flavor symmetry $SU(2)_x\times U(1)_{f_1}\times U(1)_{f_2}$ and the corresponding refined Hilbert series. 
Nevertheless, the two models have different $P$-matrices and different expressions for the generators of the mesonic moduli spaces
in terms of bifundamental chiral fields. 
These differences make explicit how distinct interaction terms can give rise to different gauge theory realizations of the same mesonic moduli space.
Furthermore, the geometric prescription yields the same modified elliptic genus for the two models.

We have also related this example to quiver-invariant duality between brane tilings through orbifold reduction \cite{Franco:2016fxm}. 
Starting from two parent $4d$ $\mathcal{N}=1$ theories with identical quivers but different superpotentials as first studied in \cite{Kho:2026geo}, 
we obtain Models A and B using a common perfect matching and the same $2$-layer sign vector $s=(-1,+1)$.
The orbifold reduction gives rise to identical daughter theory quivers, 
whereas the distinct parent superpotentials lead to different $J$-terms.
The periodic quivers also exhibit the tilting mechanism inherited from the parent brane tilings, providing a
geometric interpretation of quiver-invariant triality for the pair of models studied in this work.
Accordingly, our construction provides a link between quiver-invariant Seiberg duality in brane tilings as studied in \cite{Kho:2026geo,Lee:2026tju} and 
quiver-invariant triality in brane brick models.

A natural direction for future work is to determine more general conditions under which quiver-invariant pairs of brane tilings give
rise to quiver-invariant triality under orbifold reduction. 
It would also be useful to characterize the corresponding triality sequences directly in terms of local moves on periodic quivers and
to search for further examples in triality networks associated with a broader class of brane brick models
\cite{Bao:2024ygr,Ghim:2025zhs,Ghim:2024asj,Franco:2022gvl,Franco:2022isw,Franco:2023tyf,Carcamo:2025shw,Kwon:2026stu}.

Another direction is to investigate possible connections between quiver-invariant triality in brane brick models and other developments
in the study of brane tilings, including specular duality \cite{Hanany:2012vc}, birational transformations \cite{higashitani2022deformations,Cremonesi:2023psg,Kho:2025fmp,Kho:2025jxk},
and brane tilings associated with generalized toric polygons (GTPs) \cite{Bourget:2023wlb,CarrenoBolla:2024fxy,Franco:2023flw,Arias-Tamargo:2024fjt,Kho:2026zwc}.
Further directions include exploring connections with dimer integrable systems \cite{Goncharov:2011hp,Franco:2011sz,Eager:2011dp,Kho:2025jxk,Kho:2025fmp,Kho:2026zwc},
bipartite field theories (BFTs) \cite{Franco:2012mm,Franco:2012wv,Kho:2024wcw},
and applications of machine learning to brane tilings \cite{Krefl:2017yox,Bao:2020nbi,Choi:2023rqg,Seong:2023njx,Seong:2024wkt,Lee:2026avt}.
We leave these investigations to future work.
\\

\acknowledgments

D. G. is supported by IBS under the project code IBS-R018-D1. He is grateful to UNIST for hospitality during the various stages of this work.
R.-K. S. is supported by an Outstanding Young Scientist Grant (RS-2025-00516583) of the National Research Foundation of Korea (NRF).
He is also partly supported by the BK21 Program (``Next Generation Education Program for Mathematical Sciences'', 4299990414089) funded by the Ministry of Education in Korea and the National Research Foundation of Korea (NRF).
\textit{The authors used AI tools (OpenAI ChatGPT and Anthropic Claude) to assist with checking computations and editing the manuscript. They take full responsibility for the content of this paper.}
\\


\bibliographystyle{JHEP}
\bibliography{mybib}


\end{document}